\documentclass[8.5pt,twoside,twocolumn]{article}
\usepackage[super,sort&compress,comma]{natbib}
\usepackage{mhchem}
\usepackage{footnote}
\usepackage{hyperref}
\usepackage{times,mathptmx}
\usepackage{sectsty}
\usepackage{balance} 

\usepackage{graphicx} 
\usepackage{lastpage}
\usepackage{fixltx2e}
\usepackage{amsmath}
\usepackage{bm}
\usepackage[usenames]{xcolor}
\usepackage[format=plain,singlelinecheck=false,font=small,labelfont=bf,
labelsep=space]{caption} 
\usepackage{fancyhdr}
\definecolor{light-gray}{gray}{0.5}

\begin{document}

\thispagestyle{plain}
\fancypagestyle{plain}{
\fancyhead[L]{\textit{Submitted for consideration to RSC Molecular Biosystems}}
\renewcommand{\headrulewidth}{1pt}}
\renewcommand{\thefootnote}{\fnsymbol{footnote}}
\renewcommand\footnoterule{\vspace*{1pt}%
\hrule width 3.4in height 0.4pt \vspace*{5pt}} 
\setcounter{secnumdepth}{5}

\makeatletter 
\def\subsubsection{\@startsection{subsubsection}{3}{10pt}{-1.25ex plus -1ex minus -.1ex}{0ex plus 0ex}{\normalsize\bf}} 
\def\paragraph{\@startsection{paragraph}{4}{10pt}{-1.25ex plus -1ex minus -.1ex}{0ex plus 0ex}{\normalsize\textit}} 
\renewcommand\@biblabel[1]{#1}            
\renewcommand\@makefntext[1]%
{\noindent\makebox[0pt][r]{\@thefnmark\,}#1}
\makeatother 
\renewcommand{\figurename}{\small{Fig.}~}
\sectionfont{\large}
\subsectionfont{\normalsize} 

\fancyfoot{}
\fancyhead{}
\renewcommand{\headrulewidth}{1pt} 
\renewcommand{\footrulewidth}{1pt}
\setlength{\arrayrulewidth}{1pt}
\setlength{\columnsep}{6.5mm}
\setlength\bibsep{1pt}

\twocolumn[
  \begin{@twocolumnfalse}
\noindent\LARGE{\textbf{Uncovering allosteric pathways in caspase-1 with Markov
transient analysis and multiscale community detection}$^\dag$}
\vspace{0.6cm}

\noindent\large{\textbf{B Amor$^{\ast \, \it {a,b}}$, S N
Yaliraki\textit{$^{a,b}$}, R Woscholski\textit{$^{a,b}$} and M Barahona$^{\ast
\, \it {a,c}}$}}\vspace{0.5cm}




\noindent \normalsize{Allosteric regulation at distant sites is central to many
cellular processes.  In particular, allosteric sites in proteins are a major
target to increase the range and selectivity of new drugs, and there is a need
for methods capable of identifying intra-molecular signalling pathways leading
to allosteric effects.  Here, we use an atomistic graph-theoretical approach
that exploits Markov transients to extract such pathways and exemplify our
results in an important allosteric protein, caspase-1.  Firstly, we use Markov
Stability community detection to perform a multiscale analysis of the structure
of caspase-1 which reveals that the active conformation has a weaker, less
compartmentalised large-scale structure as compared to the inactive
conformation, resulting in greater intra-protein coherence and signal
propagation. We also carry out a full computational point mutagenesis and
identify that only a few residues are critical to such structural coherence.
Secondly, we characterise explicitly the transients of random walks originating
at the active site and predict the location of a known allosteric site in this
protein quantifying the contribution of individual bonds to the communication
pathway between the active and allosteric sites.  Several of the bonds we find
have been shown experimentally to be functionally critical, but we also predict
a number of as yet unidentified bonds which may contribute to the pathway.  Our
approach offers a computationally inexpensive method for the identification of
allosteric sites and communication pathways in proteins using a fully atomistic
description.
\\

}
\vspace{0.5cm} 
\end{@twocolumnfalse}
  ]

\section{Introduction}

\footnotetext{\dag~Electronic Supplementary Information (ESI) available:
See DOI: 10.1039/c4mb00088a}
\footnotetext{\textit{$^{a}$~Institute of Chemical Biology, Imperial College
London, South Kensington Campus, London, SW7 2AZ, UK.}}
\footnotetext{\textit{$^{b}$~Department of Chemistry, Imperial College London,
South Kensington Campus, London, SW7 2AZ, UK.}}
\footnotetext{\textit{$^{c}$~Department of Mathematics, Imperial College London,
South Kensington Campus, London, SW7 2AZ, UK.}}
\footnotetext{\textit{$^{*}$~ email: b.amor11@imperial.ac.uk, m.barahona@imperial.ac.uk}}
Allostery describes the widely observed phenomenon by which a perturbation at
one site of a protein has a functional effect at another, distant
site\cite{cui2008allostery}.  Traditionally, studies of allostery have been
linked to the cooperativity observed in large multimeric proteins such as
haemoglobin.  In this context, the classic `induced-fit'
(Koshland-Nemethy-Filmer, KNF)\cite{koshland1966comparison} and `pre-existing
equilibrium' (Monod-Wyman-Changeaux, MWC)\cite{monod1965nature} models both
consider that each monomer has a high-affinity (`relaxed') R-state and a lower
affinity (`tense') T-state.  In the induced-fit model, binding of one subunit
\textit{drives} the next subunit into its new higher affinity R-conformation. 
In the MWC model, the protein ensemble is already in equilibrium between the T
and R states, and binding of the ligand to one subunit \textit{shifts} the
equilibrium of the ensemble towards the high-affinity state. 

The so-called `new' view of allostery, which is essentially an extension of the
MWC model to general allosteric effects, regards the allosteric effector as
shifting the equilibrium of a pre-existing ensemble of conformations towards the
less populated state~\cite{Gierasch2006}.  In this sense, any protein could  be
allosteric~\cite{Gunasekaran2004}, and the perturbation could be anything that
changes the free-energy landscape of the protein (including perturbations which
do not induce a visible conformational
change\cite{cooper1984allostery,popovych2006dynamically}).  This perspective
views proteins as highly dynamic, with the ability to sample their active (i.e.,
less populated) state even in the absence of a ligand or
substrate\cite{henzlerwildman2007}.

These thermodynamic models provide a helpful phenomenological description but do
not explain how the change in the energy landscape is induced by the effector,
and how this effect physically propagates between the effector site and the
active site\cite{cui2008allostery,zhuravlev2010protein}.  The idea of allosteric
pathways (i.e., sets of contiguous residues through which a signal
propagates\cite{del2009origin}) has grown in popularity since Lockless and
Ranganathan identified a set of evolutionarily conserved residues linking the
binding site and a distal site in the PDZ family of
proteins\cite{lockless1999evolutionarily}. 
NMR\cite{fuentes2004ligand,fuentes2006evaluation} and molecular
dynamics\cite{kong2009signaling,ota2005intramolecular} studies on members of
this family have suggested other overlapping pathways of energetically linked
residues connecting the binding site to further distal sites.  Further
computational studies have used combinatorial unfolding of protein structures
combined with free energy calculations to find distal regions of proteins which
are `energetically coupled'\cite{hilser2006statistical,pan2000binding}, but do
not give a structural interpretation of this coupling.

Recently, residue-residue interaction networks (RRINs) have been used to model
allosteric communication. Using such coarse-grained representations, Del Sol
\textit{et al} identified central residues that contribute most to reducing the
average shortest path in the RRIN network~\cite{del2006residues}, and showed
that central residues often lie at the interface of modules in the
RRIN~\cite{del2007modular}. These studies used classical extremal network
measures such as betweenness to identify central residues, i.e., these measures
are based on shortest path calculations. However, allosteric communication is
characterised by multiple major and minor pathways\cite{del2009origin}.  Other
studies have used \textit{ad hoc} deterministic approaches to consider these,
such as including pathways within some small distance of the shortest
path~\cite{sethi2009dynamical,ghosh2007study}.  Taking a stochastic perspective,
Chennubhotla and Bahar~\cite{chennubhotla2007signal} (followed by
others~\cite{park2011modeling,lu2009perturbation}) used hitting and commute
times of random walks on RRINs to explore intra-protein communication and linked
these measures to the equilibrium fluctuations of a  coarse-grained,
residue-based Gaussian Network Model of the protein
\cite{chennubhotla2007signal}. 
Stochastic methods 
\cite{kong2009signaling,
chennubhotla2006markov} have also been used to analyse clustering in protein
structures.  

In this paper, we study the pathways implicated in allosteric regulation by
considering the transients of random walks taking place on an atomistic graph
representation of the protein. Our approach differs from the above methods in
two crucial ways.  Firstly, we use an atomistic rather than a residue level
description, i.e., each node in our graph corresponds to an atom (rather than a
residue) with edge weights corresponding to the actual strengths of the bonds
between atoms.  This allows us to quantify the contribution of specific atomic
interactions to communication pathways.  Secondly, we use a dynamics-based
\textit{multiscale} method, Markov Stability\cite{delvenne2010stability}, to
analyse the community structure in the protein graph across all scales in one
sway, from chemical groups to protein subunits.  In contrast, other community
detection methods (such as Modularity\cite{newman2006modularity,del2007modular})
find just one partition at a particular scale\cite{fortunato2007resolution}, and
suffer from a resolution limit~\cite{schaub2012markov} that prevents them from
identifying the multilayered community structure that exists at different scales
in highly organised networks such as proteins.  Our method overcomes this
limitation because it scans across \textit{all} scales identifying the levels of
resolution where there is strong community structure. This is equivalent to the
observation of a Markov transient over different time scales.  Such transient
behaviour can be explicitly used to explore the communication between specific
sites in the graph. To do so, we introduce a measure for intra-protein
communication, the characteristic transient time $t_{1/2}$, which takes into
account  \textit{all} possible pathways between the communicating sites. We then
use $t_{1/2}$  to identify groups of atoms that are strongly linked to the
active site, as well as bonds that are key participants in the communication
paths. In contrast to shortest path methods, our approach considers the
contribution of all possible pathways between the two sites.  Furthermore, the
computational efficiency of our method enables us to carry out full mutational
analyses to evaluate the relevance of all bonds and residues in the structure.

As a case study, we consider here the cysteine protease \textit{caspase-1}, an
important enzyme in which extensive experiments have identified an allosteric
site\cite{scheer2006common} and details of a communication pathway between the
allosteric and active sites\cite{datta2008allosteric}.
Caspase-1 processes the pro-inflammatory cytokine interleukin-1$\beta$ and has
specificity for substrates with aspartic acid adjacent to the peptide bond being
broken\cite{sleath1990substrate}.    Members of the caspase family are involved
in signalling pathways associated with apoptosis and inflammatory response, and
as such are promising drug targets\cite{li2008caspases}.  

\begin{figure}[!t]
\centering
 \includegraphics[]{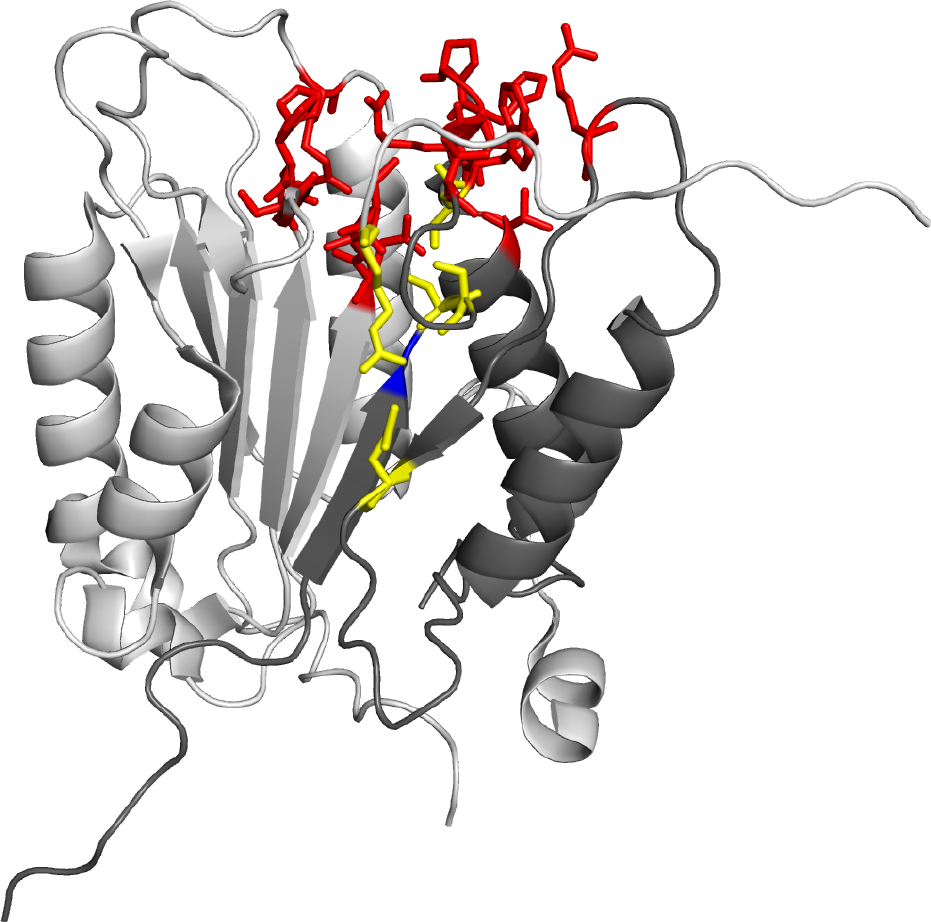}
	  \caption{Three-dimensional structure of the active conformation of
caspase-1 (PDB: 2HBQ) showing the active site residues (in red), the allosteric
binding site residue Cys331 (in blue), and the residues involved in the hydrogen
bonding network (yellow sticks)\cite{datta2008allosteric}.  The structure is a
dimer formed by the large p20 subunit (light grey) and the smaller p10 subunit
(dark grey).  All protein figures in this paper were created with PyMol
(http://www.pymol.org).}
  \label{fgr:example}
\end{figure}

Caspase-1 is a dimer composed of a smaller p10 subunit and a larger p20
subunit\cite{wilson1994structure}, with experimental indications that two such
dimers might combine to form a (p20)\textsubscript{2}/(p10)\textsubscript{2}
tetramer. Here we consider the dimer structure, which has been well
characterised structurally.  The active site spans across both subunits: 
residue Cys285 in the p20 subunit is the active site nucleophile, while the rim
of the binding pocket is composed of residues 283-285 and 236-238 in the larger
p20 subunit and residues 338-343 in the smaller p10 subunit.  Scheer \textit{et
al} have identified experimentally an allosteric binding pocket situated at the
dimer-dimer interface\cite{scheer2006common}. Datta \textit{et
al}\cite{datta2008allosteric} used structural data to identify a network of
hydrogen bonds which link the allosteric and active sites in the active
conformation (Fig.~1), but are absent in the inactive conformation. A subset of
those residues were then found to have a significant impact on catalytic
activity.  In particular, binding of the allosteric ligand disrupts a
salt-bridge between residues Arg286 and Glu390, and an alanine mutation of
either residue greatly reduces the catalytic activity of the protein.

Our analysis of the active and inactive conformations of caspase-1 shows that
the active conformation has a less compartmentalised community structure than
the inactive structure, which leads to increased communication between the
active site and the rest of the protein. Through computational exploration of
all point mutations, we find that only a few residues have a strong impact on
this increased communication. We then consider the explicit analysis of
transient random walks in the structures and find that the region of the protein
with the largest increased connectivity in the active conformation corresponds
to the allosteric site. Our computational mutational analysis then shows that
the bonds that contribute most to this increased connectivity are formed by
functionally important residues. Our method thus reveals the location of the
allosteric site and the bonds involved in signal transmission directly from
structural data.

\section{Materials and methods}

\subsection{Structural data}

We analyse three crystal structures of human caspase-1: one in unliganded
(`inactive') form, and two complexed (`active') structures  in complex with
tetrapeptide substrates at the active site.
The unliganded structure (PDB ID code: 1SC1) was obtained by Romanowski
\textit{et al}\cite{romanowski2004crystal} through X-ray crystallography at a
resolution of 2.6$\AA$.  
The complexed caspase-1/z-VAD-FMK (PDB ID code: 2HBQ) was obtained by Scheer
\textit{et al}\cite{scheer2006common} at a resolution of 1.8$\AA$.  
The structure of caspase-1 in complex with a tetrapeptide aldehyde inhibitor
(PDB ID code: 1ICE) was obtained by Wilson \textit{et
al}~\cite{wilson1994structure} at a resolution of 2.6$\AA$.  We follow the
standard residue numbering as in Ref.~\citet{wilson1994structure}. 

\subsection{Construction of the atomistic network}

In contrast to most network methods for protein
analysis~\cite{park2011modeling,lu2009perturbation,chennubhotla2006markov,
chennubhotla2007signal,del2007modular,del2006residues}, our method starts by the
construction of a fully atomistic graph representation of the protein.  
The graph is built from the structural information contained in the PDB
file\cite{berman2000protein}, which contains the Cartesian coordinates of each
atom in the protein.  Each node in the graph corresponds to a single atom and
each edge defines a covalent bond or weak interaction (hydrogen bonds, salt
bridges, or hydrophobic tethers)~\cite{delmotte2011protein}.  Any missing
hydrogen atoms are added using the software Reduce~\cite{word1999asparagine}. We
identify the presence of covalent bonds and weak interactions using the program
FIRST\cite{jacobs2001protein} with a cutoff of 0.01 kcal/mol for hydrogen bonds
and 8\AA\ for hydrophobic interactions. Each edge has a weight which is linearly
related to the bond energy. The bond strengths are obtained from the DRIEDING
force-field\cite{mayo1990dreiding}. This protein graph is encoded by a weighted
adjacency matrix $A$, an $N \times N$ symmetric matrix (where $N$ is the number
of atoms in the protein) in which the entry $A_{ij}$ gives the energy of the
interaction between atoms $i$ and $j$ (0 if there is no interaction).    For
further details of the network construction, including the force fields used,
see the Supplementary Information and Ref.~\citet{delmotte2011protein}.

\subsection{Multiscale community detection with Markov Stability for protein
structures}

\textbf{Intuition for Markov Stability analysis of atomic protein graph.} 
Proteins are multiscale biomolecular machines with structural organisation at
scales ranging from chemical groups, through amino acids, to protein domains,
and even different subunits.  This multiscale structural organisation is encoded
in the atomic protein graph described above, which contains detailed
physico-chemical and geometric properties of the protein. To reveal the
organisation of the structure at different scales, we analyse the generated
protein network using recently developed graph-theoretical techniques. 

Markov Stability is a general method for multiscale community detection in
graphs~\cite{delvenne2010stability}, and is thus well suited for the analysis of
the protein graph at all scales. In contrast to other community detection
methods (such as Modularity\cite{newman2006modularity}), Markov Stability finds
partitions of a graph into non-overlapping subgraphs (`communities') without
imposing a particular scale \textit{a priori}. Rather than obtaining a single
partition, Markov Stability finds an optimised partition at every scale and
decides if such a partition is significantly robust by establishing if the
communities correspond to subgraphs where a random walk is likely to remain
trapped over a certain timescale. As we increase this Markov timescale, the
method acts as a zooming lens, scanning across all scales looking for
significant communities at different resolutions. Hence, as the Markov time
progresses, the partitions become coarser. In the case of the protein graph,
this process allows us to scan across resolutions and find communities involving
a few atoms (corresponding to chemical groups) at very short Markov times,
through biochemical units (amino acids) and secondary structures (helical turns)
at intermediate Markov times, to communities corresponding to conformational
groupings or protein subunits at long Markov times. For a detailed description
of the method, see the Supplementary Information and
Refs.~\citet{delvenne2010stability,delmotte2011protein,schaub2012markov}.

\textbf{Optimal partition of a graph into communities at a given scale.}  
We now make these notions more precise.
A random walk on a graph is described by a $N \times N$ Markov transition matrix
$M$, where $N$ is the number of nodes in the graph:
\begin{equation}
\mathbf{p}_{t+1} = \mathbf{p}_t \, M.
\label{eq:RW}
\end{equation}
Each element $M_{ij}$ gives the probability of transitioning from node $i$ to
node $j$ in one time step and $\mathbf{p}_t = (p_t^{(1)},\ldots,p_t^{(N)})$ is a
$1 \times N$ probability vector recording the probability of the process at each
node at time $t$.  $M$ is directly related to the adjacency matrix by $M =
D^{-1}A$, where $D$ is the diagonal matrix of node degrees, i.e., $D_{ii}$ is
the sum of the weights of all edges incident to node $i$.  

\begin{figure}[!t]
\centering
  \includegraphics[]{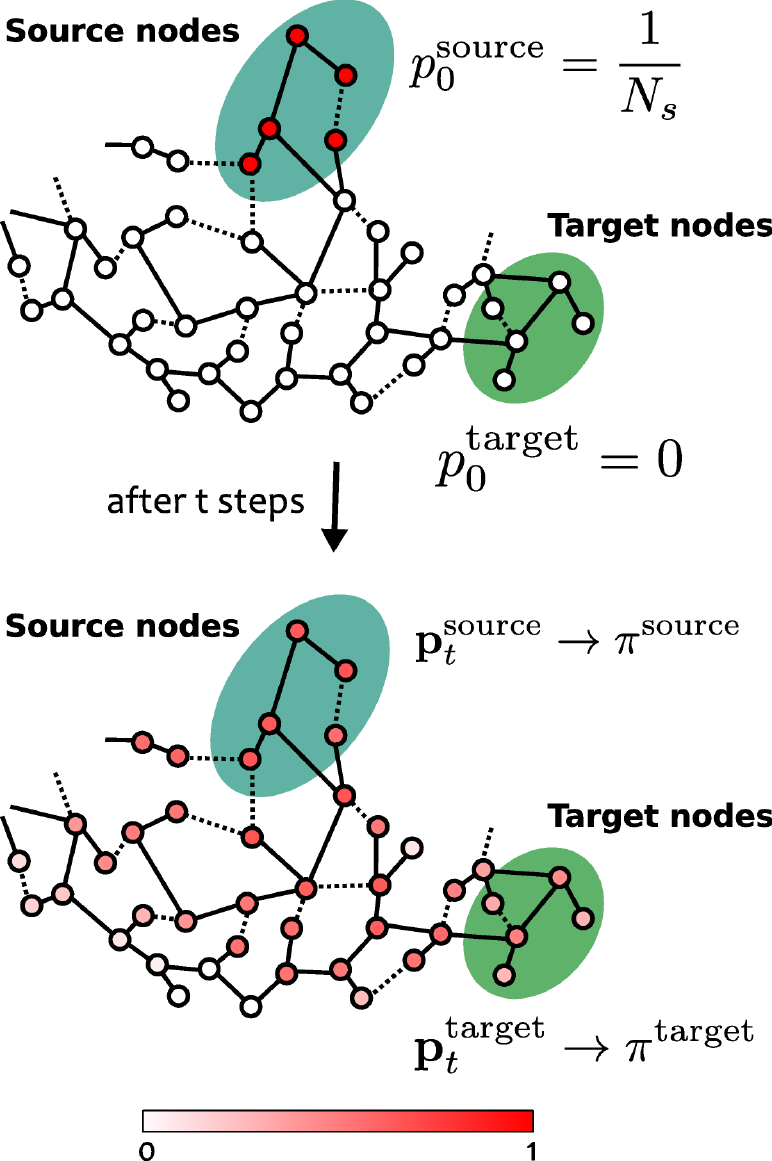}
	  \caption{\textbf{Schematic illustration of random walk analysis}. 
Initially the probability is distributed only across the source nodes.  Through
the iteration of the Markov process~\eqref{eq:RW},
the probability spreads across the other nodes in the protein.  We monitor the
time evolution of the probability at a set of target nodes and measure how long
it takes for this set to reach half their stationary value.}
  \label{fgr:example}
\end{figure}

The key matrix for Markov stability is the block `autocovariance'
matrix\cite{delvenne2010stability}
\begin{equation}
R(t,H) = H^{T}(\Pi M^{t} - \pi^{T}\pi)H.
\end{equation}
$R$ is a $c \times c$ matrix where $c$ is the number of communities in the
partition, and $[R(t)]_{ij}$ gives the probability of the random walker starting
in community $i$ and finishing in community $j$ after $t$ timesteps.  The $N
\times c$ indicator matrix $H$ encodes the membership in the partition, while
the $1 \times N$ vector $\pi = (\pi^{(1)},\ldots,\pi^{(N)})$ is the stationary
distribution of the random walk~\eqref{eq:RW} and $\Pi = \text{diag}(\pi)$.  

We look for partitions where a random walker is likely to remain trapped in the
same community over the timescale $t$. This corresponds to large values of the
diagonal elements $[R(t)]_{ii}$.   
Therefore, the Markov Stability of a partition $H$ is defined as
\begin{equation}
r(t) = \max_{H} \text{trace}(R(t,H)),
\end{equation}
and we search for partitions $H$ that maximise $r(t)$.   
The `time' $t$ is a dimensionless quantity that measures the expansion of the
random walk in the network acting as a dynamical resolution parameter, and does
not directly correspond to a biophysical timespan. To emphasise this difference,
we refer to $t$ as the \textit{Markov time} throughout. To optimise the Markov
Stability $r(t)$ for any given $t$, we use the Louvain
algorithm~\cite{blondel2008fast},  
a highly efficient greedy algorithm that finds optimised partitions with high
values of $r(t)$ with no
guarantees of global optimality (the problem is NP-hard), but which works well
in practice.

\textbf{Robustness as a measure of significant graph partitions.}  At any Markov
time $t$, the above process finds an optimised partition but there is no
\textit{a priori} reason why this partition need have any significance.  Indeed,
it is likely that significant partitions will only be found at certain scales
(e.g., at the level of chemical groups or of amino acids). Through the use of
surrogate chemical randomisations \cite{delmotte2011protein}, we have shown that
Markov Stability is able to detect chemical groups, biochemical units, as well
as structural features such as helical turns.  At long Markov times, we look for
significant partitions with the defining feature of being robust to
perturbations\cite{karrer2008robustness}.  We use two measures to quantify this
robustness:
\begin{enumerate}
\item The length of the Markov time over which the partition is found optimal is
an indication of its persistence for flow retention under small parametric
changes in time.  Hence we look for plateaux in the number of communities versus
Markov time.
\item A widespread similarity between the partitions found by the optimisation
algorithm indicates the robustness of the partition found.   At each Markov
time, we obtain 100 optimised partitions using 100 different initial conditions
for the optimisation algorithm.  We then calculate the average pairwise
difference between these 100 partitions using the \textit{variation of
information} (VI)\cite{meilua2007comparing}, an information-theoretic measure
that quantifies the similarity/dissimilarity of two partitions of the same
network. A low VI (or a dip in the VI) reflects greater homogeneity among the
100 optimised solutions obtained and therefore increased robustness to the
optimisation.
\end{enumerate}

\textbf{\textit{In silico} mutational analysis to identify significant
residues.}  Alanine scanning mutagenesis is the systematic experimental
replacement of individual amino acids with alanine in the primary structure of a
protein.  We can mimic computationally the effect of this procedure by removing
the graph edges corresponding to the weak interactions formed by the side chain
of the chosen amino acid.  The computational efficiency of our algorithm means
we can evaluate the effect on the Markov Stability partitions of all individual
mutations of each residue in turn.  To identify the Markov times at which a
mutation has a significant impact on the robustness of a partition, we compare
the VI($t$) graphs of the wild type versus that of the mutated networks. 
Gaussian Process Regression\cite{rasmussen2006gaussian} is used to obtain a
representative VI curve for the ensemble of mutated and wild-type VI graphs.  If
the VI graph of a mutated network falls outside the statistical bounds of the
ensemble trajectory, it indicates that the robustness of the partition has been
significantly affected (Fig.~S2). We identify the point mutations that lead to
significant changes in the structurally relevant graph partitions.  See section
S3 for a full discussion of Gaussian Process Regression.   

\subsection{Markov transient analysis and signal propagation}
To identify special regions in the protein that are significantly connected to
the active site we perform an explicit analysis of Markov transients from
initial conditions localised on a particular subgraph and establish a measure of
the convergence of another target subgraph towards stationarity.

\textbf{Source-target transients of the random walk.}  
Consider the evolution of a random walk described by Eq.~\eqref{eq:RW}.  To
model the propagation through the network of a perturbation occurring at a
particular site, we analyse a random walk originating at that site and define an
initial probability distribution $\mathbf{p}_{0}$ in which the probability is
spread uniformly over a set of \textit{source nodes}:
\begin{equation*}
\begin{array}{ccccccc}
& & \quad \mathbf{p}_{0}^\text{source} & & \quad \mathbf{p}_{0}^\text{target} &
\\
\mathbf{p}_{0} = \Big( 0 & \ldots &  
\overbrace{
\begin{bmatrix}
\frac{1}{N_s} 
\ldots
\frac{1}{N_s} 
\end{bmatrix}
}
& \ldots & 
\overbrace{
\begin{bmatrix}
0 \ldots 0
\end{bmatrix}
}
& \ldots & 0 \Big ),
\end{array}
\end{equation*}
where $N_s$ is the number of source nodes.  
We then monitor signal propagation between two defined regions of the network by
observing the change in probability at the \textit{target nodes}:
\begin{equation}
\mathbf{p}_{t} =
\begin{pmatrix} 
\ldots 
\begin{bmatrix}
\mathbf{p}_{t}^{\text{source}}
\end{bmatrix}
\ldots 
\begin{bmatrix}
\mathbf{p}_{t}^{\text{target}}
\end{bmatrix}
\ldots
\end{pmatrix},
\end{equation}
As the Markov time $t \rightarrow \infty$, the vector $\mathbf{p}_{t}$ 
converges to the stationary distribution $\pi$. Hence the speed at which the
target nodes reach stationarity can be used as a measure of connectivity between
the source and target nodes. See Fig.~2 for a sketch of this procedure.

\begin{figure}[!t]
\centering
  \includegraphics[]{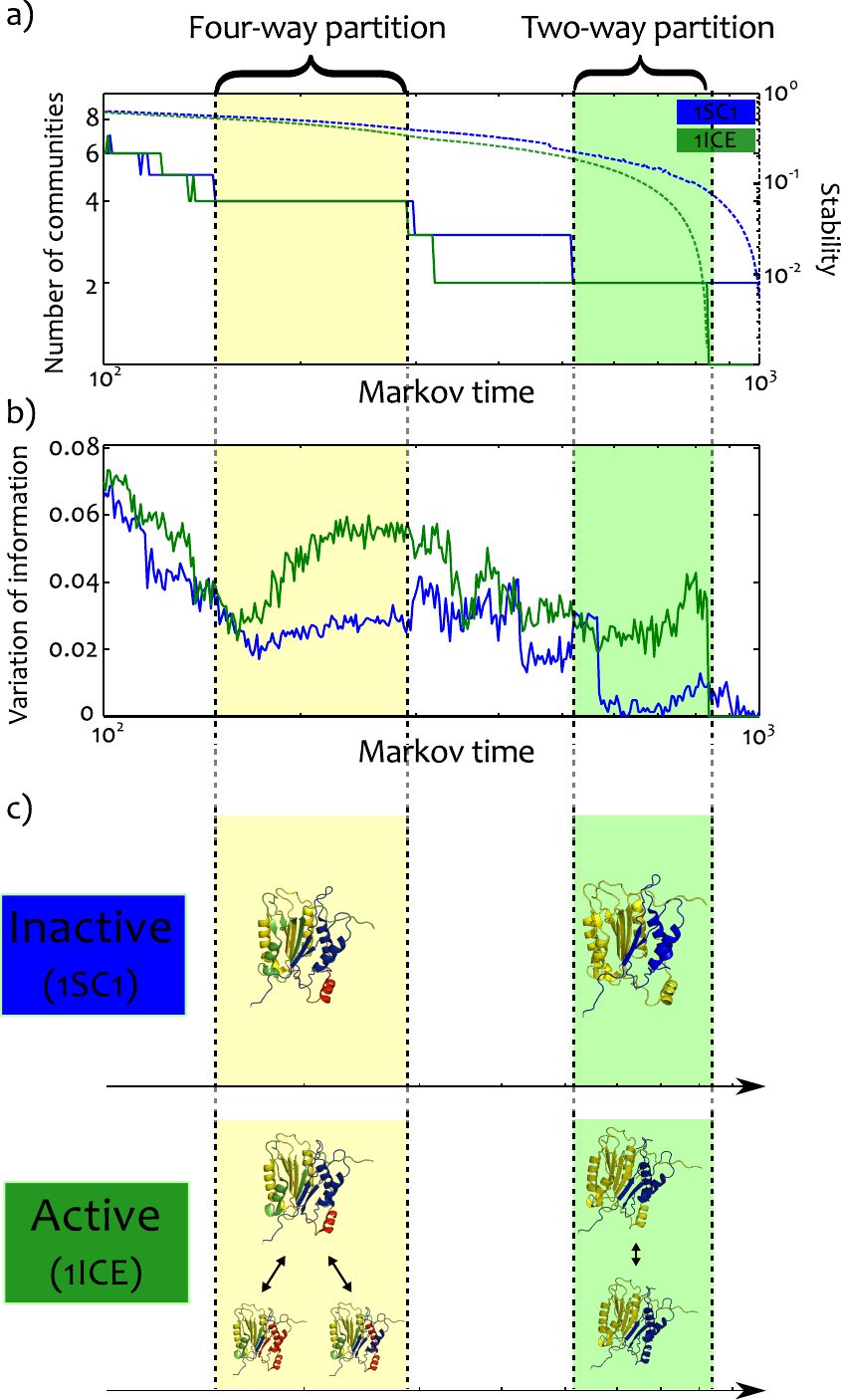}
  \caption{\textbf{Community detection across timescales for caspase-1.} a)
Number of communities (solid line) and Markov Stability (dashed line) between
Markov times 100 and 1000 of two conformations of caspase-1: active (green) and
inactive (blue). b) The variation of information of both conformations over the
same time scale indicates that the 4-way and 2-way partitions of the active
conformation are less robust than for the inactive conformation. c) The dominant
4-way and 2-way partitions for the two conformations.  The inactive conformation
has a single dominant partition, the inactive conformation flips between several
different partitions.}
  \label{fgr:example}
\end{figure}

\textbf{Characteristic transient time $t_{1/2}$ as a measure of signal
propagation between two sites.}  To measure the connectivity between two sites
in the protein, we introduce the characteristic transient time $t_{1/2}$,  a
measure of the speed with which a random walk originating at one site will
propagate to the target site. Given a set of source atoms, the $t_{1/2}$
associated with the target node $i$ is the number of time steps it takes for the
probability at node $i$ to reach half its stationary value: 
$$t_{1/2}^{(i)} = \operatorname*{arg\,min}_{t} \left[p_t^{(i)} \geq
\frac{\pi^{(i)}}{2}\right].$$ 
To measure the connectedness between two \textit{sets} of atoms, we average the
$t_{1/2}^{(i)}$ over all atoms in the target set (e.g., over the atoms of a
residue or over the atoms of a group of residues)

\textbf{Measuring changes in signal propagation: the $t_{1/2}$ ratios.}  
We will be interested in measuring the changes in signal propagation from the
active site to any other residue (or groups of residues) in the protein as
quantified using the characteristic transient time $t_{1/2}$ defined above.  We
define three ratios that allow us to compare the $t_{1/2}$ under different
changes in the protein:
\begin{description}
\item \textbf{Conformational $t_{1/2}$ ratio:} measures the change in $t_{1/2}$
for each residue for a random walk originating at the active site when comparing
the active and inactive conformations of the protein
\begin{equation}
\label{eq:CF}
\Delta_{\text{CF}} = \frac{t_{1/2}^{\text{inactive}}}{t_{1/2}^{\text{active}}}.
\end{equation}
A high ratio indicates that a residue is more closely coupled to the active site
in the active conformation.
\item  \textbf{Bond-removal $t_{1/2}$ ratio:}  measures the contribution of
individual bonds to signal propagation (between the active site and a target) by
comparing the $t_{1/2}$ before and after removal of a bond in the graph
\begin{equation}
\label{eq:BR}
\Delta_{\text{BR}} =\frac{t_{1/2}^{\text{bond-rem}}}{t_{1/2}^{\text{active}}}.
\end{equation}
\item \textbf{Mutational $t_{1/2}$ ratio:}  measures the importance of
individual residues to communication between two sites by comparing the
$t_{1/2}$ before and after mutation of that residue (i.e. by removal of
\textit{all} weak interactions formed by that residue) 
\begin{equation}
\label{eq:MT}
\Delta_{\text{MT}} =  \frac{t_{1/2}^{\text{mut}}}{t_{1/2}^{\text{active}}}.
\end{equation}
\end{description}

\section{Results and discussion}

\subsection{Markov Stability reveals a strongly compartmental community
structure in the inactive conformation of caspase-1}
\label{sec:Markov}

We have used Markov Stability to analyse the `active' (1ICE) and `inactive'
(1SC1) structures of caspase-1. The 1CE active structure is used as it does not
have unresolved residues causing a break in the backbone chain. Note that the
ligand in 1ICE is not included in the graph, so as to make the comparisons with
the unliganded 1SC1 consistent, i.e., the observed differences between the
conformations are due to changes in graph properties induced on the protein
structures and not due to the extra atoms/bonds of the ligand. 

As described above, Markov Stability zooms across different levels of resolution
to find robust graph communities at all scales from the atomic graph of the
protein. Such communities can be thought of as groups of atoms behaving
coherently over a particular timescale under a diffusive process. The full
Markov Stability analysis is shown in Fig.~S1.  Drops in the variation of
information (VI), indicative of robust partitions, are observed around Markov
times $2 \times 10^{-3}$ and $10^{-1}$ corresponding to chemical groups and
amino acids, respectively, as discussed elsewhere~\cite{delmotte2011protein}. 
Between Markov times $3\times 10^{-1}$ and 10, high VI and a lack of plateaux
indicates an absence of significant partitions in the protein structure.  The VI
curve begins to fall again after Markov time 10, and around this time we see the
emergence of the secondary structure with alpha-helices forming communities. 
These features are observed in all proteins studied with this method so
far\cite{delmotte2011protein}, and confirm that we uncover the expected
structure for proteins at small and intermediate scales.  Consequently, for
these short and intermediate Markov time scales, there are no differences
between the active and inactive conformations of caspase-1 since they share
their chemical constituents.

\begin{figure*}[!t]
\centering
 \includegraphics[width=\textwidth]{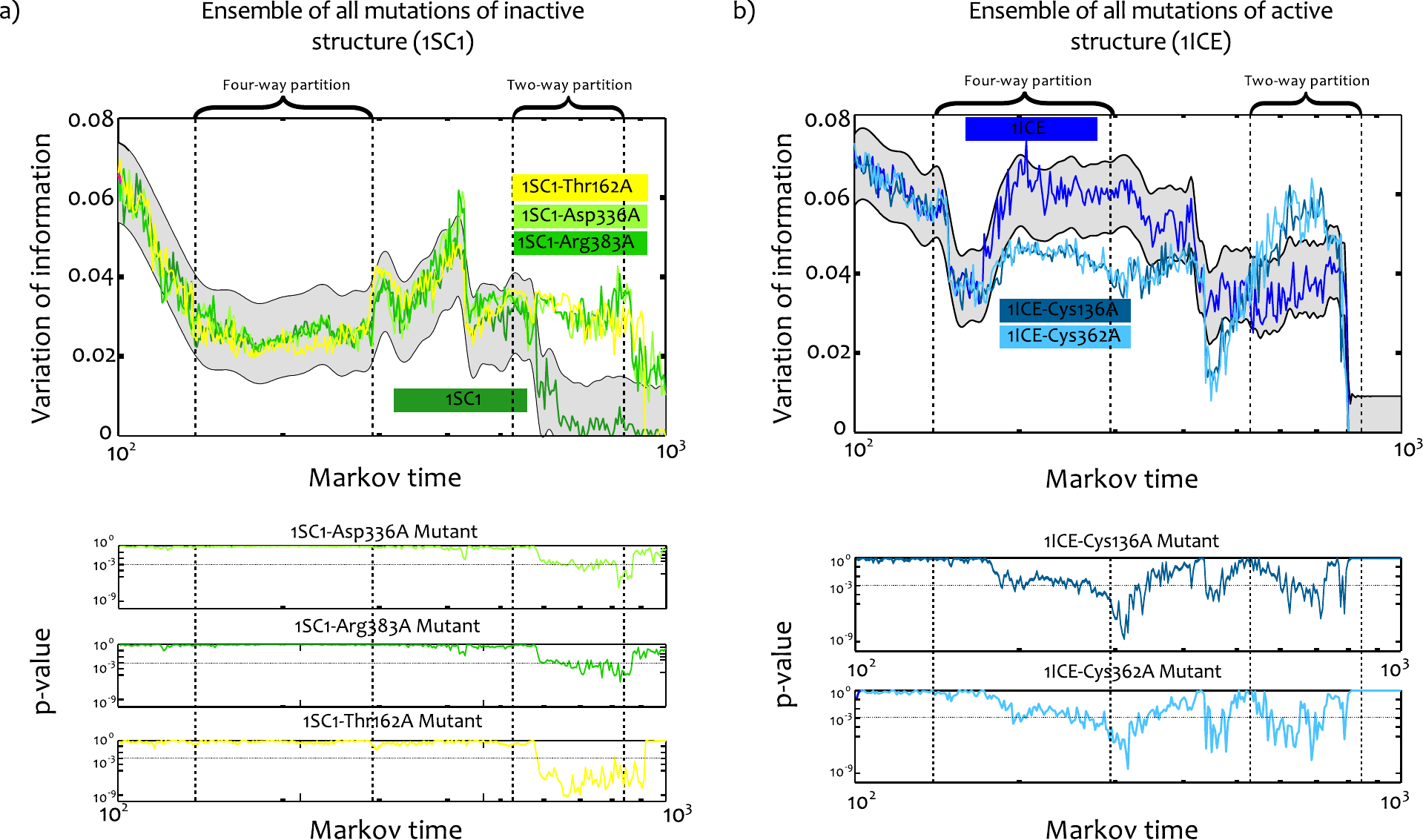}
  \caption{\textbf{Variation of information (VI) as a function of Markov time
for mutated structures} a) Analysis of all the mutated inactive structures: (top
panel) VI as a function of Markov time with statistical bounds (95\% confidence
intervals) obtained with GPR applied to the ensemble of all mutations.  Above
Markov time 570 (corresponding to the 2-way partition), the VI for mutations of
Thr162, Asp336 and Arg383 are well above the statistical bounds of the ensemble,
reflecting a loss of well-defined community structure in the two-way partition
similar to what is observed in the active structure; (bottom panel) The p-value
quantifying the likelihood that the mutated structure does not belong in the
ensemble of mutations drops sharply ($p< 0.001$) over the 2-way plateau. b)
Analysis of all the mutated active structures: the VI for mutations of Cys136A
and Cys362A drop below the statistical bounds between Markov times 170 and 400,
representing an increased robustness of the four-way partition observed during
this period (top panel). These mutations also induce changes in the robustness
of the two-way partition as is also seen in the p-values shown in the bottom
panel.   
}
\end{figure*}

Above Markov time 100, however, the results for the active and inactive
conformations diverge and we see distinct features for each conformation of
caspase-1 (Fig.~3). In particular, the inactive conformation exhibits long
plateaux in the number of communities between Markov times 120-300 and 460-800,
accompanied by a drop in the VI, indicating the presence of robust four-way and
two-way partitions (Fig.~3).  On the other hand, the active conformation
presents much weaker indications of community structure, i.e., the quantitative
Markov Stability curve of the inactive structure (which measures the strength of
the partition) is consistently smaller and the VI is larger during the
plateaux. 

In particular, the four-way partition in the inactive conformation is more
robust, with an average pairwise VI $\simeq 0.0254$ in the inactive structure
compared to the larger 0.0435 in the active structure.  In the inactive
structure there are four clearly demarcated communities which comprise the
four-way partition (Fig.~3c top left).  In contrast, in the active conformation
we identify 7 smaller sub-communities which combine in different ways to form
different four-way partitions in a more flexible manner (Fig.~3c bottom left).  

Furthermore, the long-lived, robust (VI $\simeq 0.0073$) two-way partition of
the inactive conformation splits caspase-1 into its p10 and p20 subunits
(Fig.~3c top right).  In the active conformation, however, the two subunits are
less well defined as separate communities (VI $\simeq 0.0281$): the $\alpha-1/2$
helix and the $\beta-6$ strand of the p20 subunit are closely associated with
the p10 subunit (Fig 3c. bottom right), indicating a stronger interaction
between the p10 and p20 subunits in the active conformation.

\subsection{Computational mutagenesis reveals important residues for community
structure}
To mimic \textit{in silico} the process of alanine mutagenesis, we remove all
edges corresponding to interactions of a given `mutated' residue. We consider
all point mutations in turn, and compute the community structure using Markov
Stability for each mutated structure.  We can then identify the mutations that
affect significantly the robustness of the 4-way and 2-way partitions by
analysing the VI of the ensemble of mutated networks using Gaussian Process
Regression\cite{rasmussen2006gaussian} (see Section S3).  Mutations are
classified as significant if the VI of the mutated structure lies 3 standard
deviations outside the mean for at least one third of the relevant Markov time
plateaux.  Using this criterion, we find that only three residues
(Thr162/Asp336/Arg383) in the inactive conformation and two residues
(Cys136/Cys362) in the active conformation affect the community structure
significantly at long Markov times (Fig.~4).

In the active structure, mutations of residues Cys136 and Cys362 significantly
affect the four-way partition (Fig.~4b). These residues form a disulphide bond
linking the p10 and p20 subunits in the active structure, which is absent in the
inactive structure (Fig.~S4a).  Removing this bond by our computational mutation
breaks a strong link between the two subunits and appears to stabilise the
four-way partition (Fig.~3c), resulting in the more compartmentalized community
structure characteristic of the inactive conformation.  Although experimental
mutations of either residue have not been shown to have an effect on enzyme
activity\cite{wilson1994structure}, they may affect dynamical and structural
features of the protein. 

The three mutations that have an effect on the large scale organisation of the
inactive structure are Thr162, Asp336 and Arg383.  Thr162 lies in the
$\alpha-1/2$ helix in the p20 subunit and forms hydrogen bonds with Glu223 and
Thr226 (Fig.~S4b).  Removal of these interactions has a significant effect on
the two-way partition of the inactive conformation at long Markov times
(Fig.~4a): instead of a clear two-way partition into the two subunits, we find
that the $\alpha-1/2$ helix of the p20 subunit forms a community with the p10
subunit. 

Asp336 and Arg383 form a salt bridge (Fig.~S4c) and, similarly, the removal of
this bond by mutation of either residue causes a greater association of the
$\alpha-1/2$ helix and the p10 subunit.  It has been suggested that the main
role of Asp336 is to stabilise the L3 loop (residues 332-346) containing residue
Arg341, an important residue for
substrate-recognition\cite{datta2008allosteric}.  Our results suggest that the
Asp336/Arg383 bond may play a role in stabilising the inactive conformation and
that removing this salt bridge may facilitate adoption of the active
conformation.  Romanowksi \textit{et al}\cite{romanowski2004crystal} suggest
that the primary function of the L4 loop, containing Arg383, may be to stabilise
loop L3 in the inactive conformation.  Our analysis suggests that mutation of
this residue may have a global effect on the dynamics of the enzyme.  In section
3.4 we find that this residue may play a role in active-site to active-site
co-operativity.

\subsection{Allosteric pathways uncovered by transients of random walks on the
atomistic graph of the protein}

\textbf{The conformational $t_{1/2}$ ratio $\Delta_{\text{CF}}$ reveals the
location of the allosteric site. }
To investigate signal propagation between the active site and the rest of the
protein, we consider first the behaviour of a random walk originating at the
active site.  The active site is defined as the residues containing an atom
within 4$\AA$ of the substrate and comprises residues
179/236-238/283-285/338-343/345/348/383\cite{wilson1994structure}.  We use the
2HBQ active structure for the transient analysis as this structure was used
previously to characterise an allosteric hydrogen bonding
network\cite{datta2008allosteric}. 

\begin{figure}[]
 \includegraphics[width=0.5\textwidth]{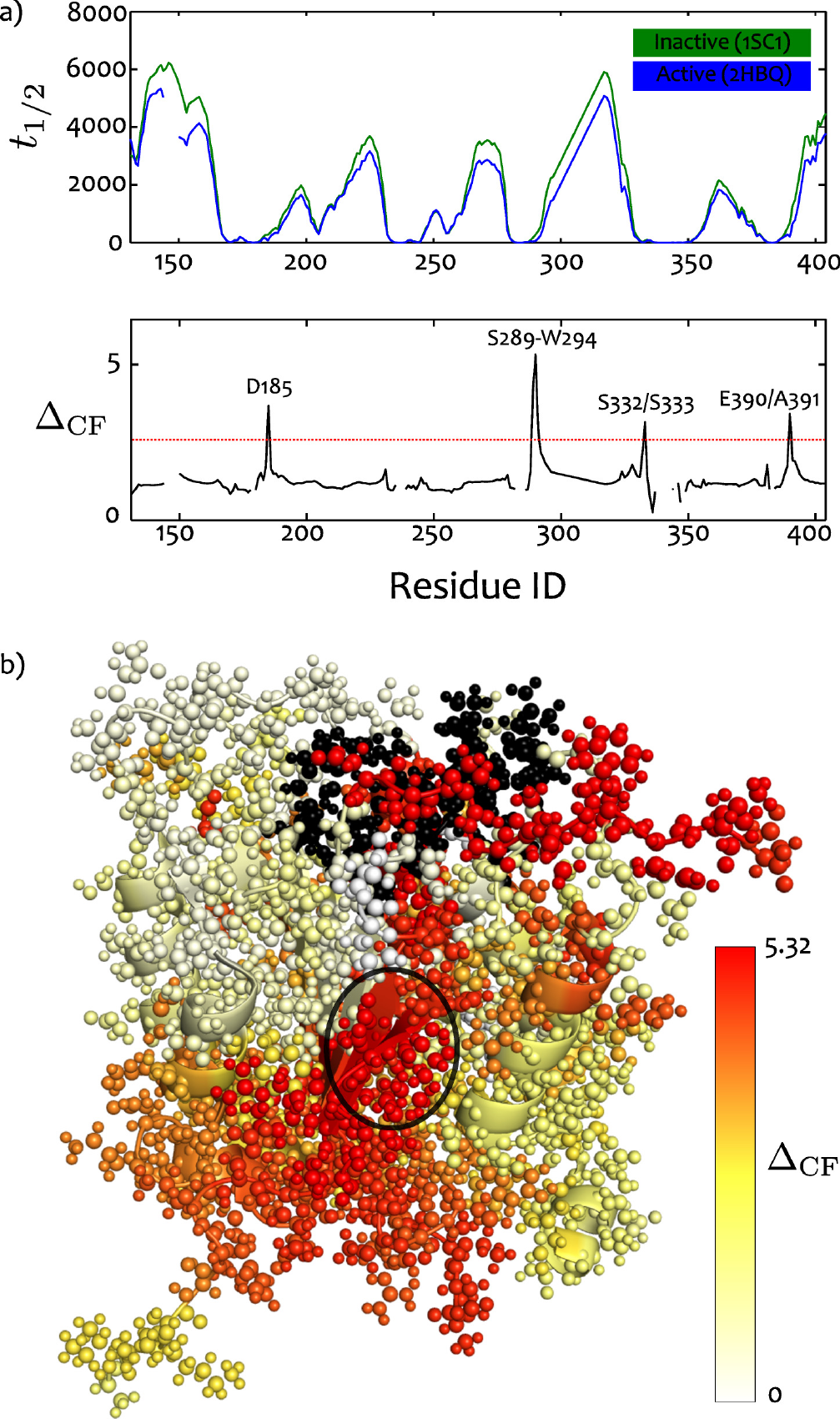}
  \caption{\textbf{Difference in signal propagation originating from the active
site between active and inactive conformations.}  a) Characteristic transient
times $t_{1/2}$ for random walks originating at the active site for all residues
for the active (green) and inactive (blue) conformations  (top panel),  and the
conformational ratio $\Delta_{\text{CF}}$ defined in Eq.~\eqref{eq:CF} (bottom
panel).  The random walk propagates more quickly in the active conformation with
$t_{1/2}^\text{inactive} > t_{1/2}^\text{active}$ consistently.  The red dashed
line corresponds to $\Delta_{\text{CF}} > 2$ and identifies the residues shown
in Table 1. b) The conformational ratio $\Delta_{\text{CF}}$ mapped onto the
protein structure: red areas show the biggest increase of $t_{1/2}$ in the
inactive conformation with a `hot spot' at the allosteric site (circled).  The
source atoms of the random walk (the active site) are coloured black.}
  \label{fgr:example}
\end{figure}

As would be expected from its weaker community structure, the random walk
spreads more rapidly through the protein in the active conformation. The average
characteristic transient time to any residue in the protein for a random walk
originating at the active site in the active conformation is ($\bar{t}^{\,\,
\text{active}}_{1/2}=1367$), much shorter than that of the inactive conformation
($\bar{t}^{\,\, \text{inactive}}_{1/2}=1663$). Crucially, the differences in
transient times between the inactive and active conformations are not
distributed homogeneously across the structure: mapping the conformational ratio
$\Delta_{\text{CF}}$ onto the protein structure reveals a `hot spot' at the
allosteric site (Fig.~\ref{fgr:example}). In Table 1, we show the ten residues
with $\Delta_{\text{CF}} > 1.9$, corresponding to the largest change between
inactive and active conformations. Three of these ten residues
(Ser332/Ser333/Glu390) are in the hydrogen bonding network by Datta \textit{et
al}\cite{datta2008allosteric}.  Of these, Glu390 is notable for being located in
the allosteric binding pocket at the dimer-dimer interface. Glu390 forms a salt
bridge with Arg286, a bond which is known to be disrupted through allosteric
inhibition.  Other residues with large conformational ratios  
$\Delta_{\text{CF}}$ are Asp185 and residues 289-293.   The large
$\Delta_{\text{CF}}$ of residues 289-293, which are located in the highly
dynamic C-terminus of the p20 subunit, is due to the loss of hydrogen bonds
between Ser289 and two active-site residues Asp336 and Val338.  Similarly,
Asp185 loses a hydrogen bond with the active site residue 179.

To test the relevance of the allosteric site residues identified, we performed
the transient analysis of the `reverse' random walk originating at the residue
Glu390 in the allosteric site.  The random walk spreads much more quickly
towards the active site in the active conformation (Fig. 6 and Video 1 in the
SI), revealing the existence of a communication pathway between the allosteric
and active sites which is present in the active conformation and suppressed in
the inactive conformation.  To identify the scaffold of this communication
pathway, we compute the conformational ratio $\Delta_{\text{CF}}$ for this
reverse random walk.  As shown in Table 2, the largest $\Delta_{\text{CF}}$
corresponds to residue Arg286, a consequence of the formation of the salt-bridge
with Glu390.  Residues 285-291, adjacent to residue Arg286 and including the
catalytic residue Cys285, also see significant increases.  Also notable is
residue Ser339 and residue Ser333, which are in the hydrogen bonding network
identified by Datta \textit{et al}.

\begin{table}
\small
  \caption{Residues with largest conformational ratio $\Delta_{\text{CF}}$ for
random walks originating at the active site.  Residues marked with an asterisk
appear in the hydrogen bonding network identified by Datta \textit{et al}
\cite{datta2008allosteric}.}
  \label{tbl:example}
  \begin{tabular*}{0.5\textwidth}{@{\extracolsep{\fill}}cccc}
    \hline
    Residue &  $t_{1/2}^\text{active}$ & $t_{1/2}^\text{inactive}$ &
$\Delta_{\text{CF}}$ \\
    \hline
    P290 & 50.57 & 268.93 & 5.32\\
    S289 & 37.55 & 154.45 & 4.11\\
    D185 & 74.50 & 273.75 & 3.67 \\
    E390\footnotemark[1] & 199.60 & 683.73 & 3.43\\
    S333\footnotemark[1] & 20.09 & 63.36 & 3.15 \\
    G291 & 132.57 & 384 & 2.90\\
    V292 & 246.25 & 538.25 & 2.19\\
    S332\footnotemark[1] & 11.18 & 23.82 & 2.13\\
    V293  & 472.25 & 950 & 2.01 \\
    A391 & 674.96 & 1297.4 & 1.92\\
    \hline
  \end{tabular*}
\end{table}

\begin{figure*}[!t]
\centering
 \includegraphics[width=\textwidth]{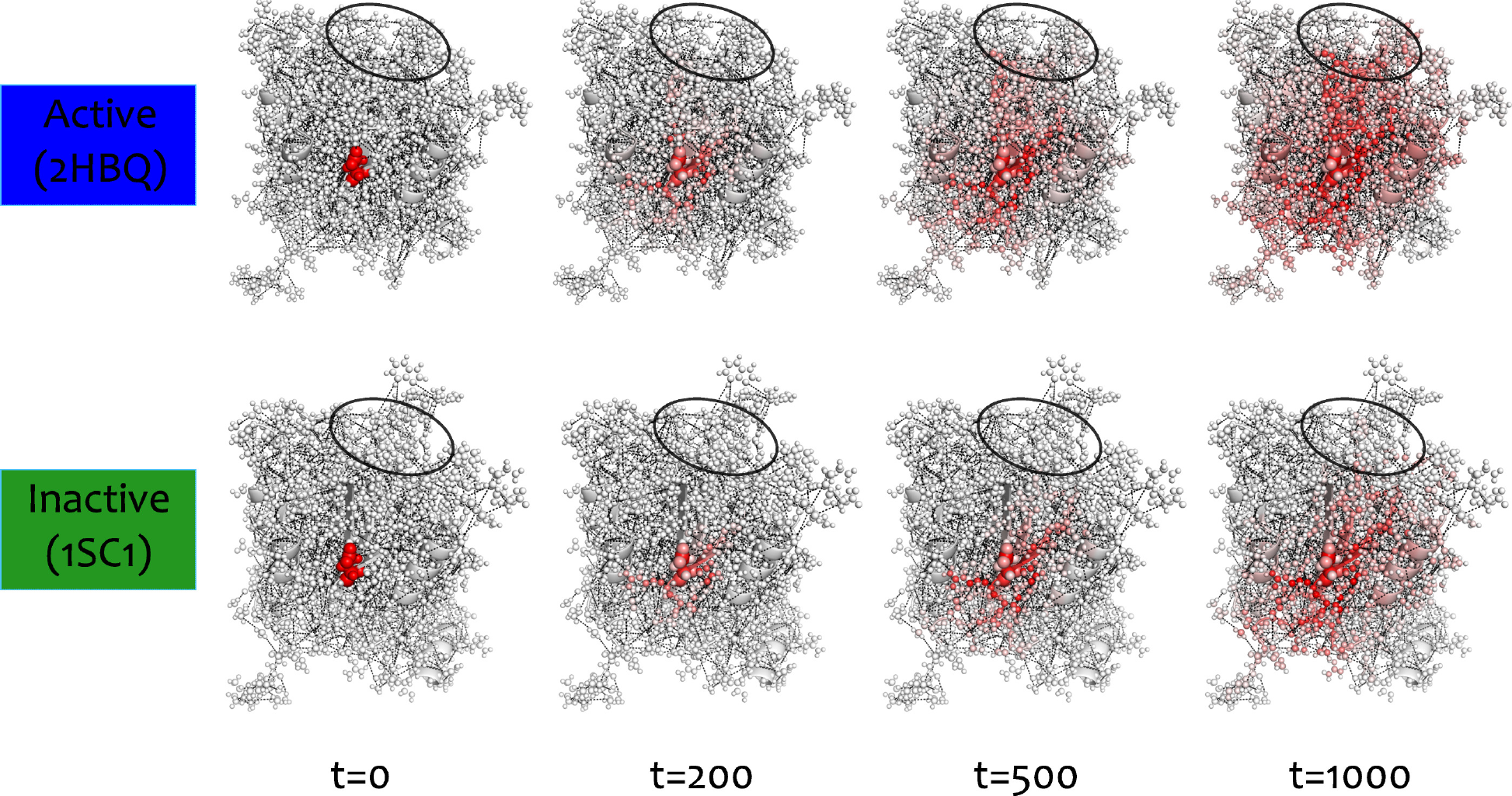}
  \caption{\textbf{Propagation of a random walk originating at Glu390.}  The
probability of the random walk originating from residue Glu390 is shown for time
steps $t=0,200,500,1000$.  Atoms are coloured from white to red with increasing
probability.  The spread of the random walk in the active structure is broader,
reflecting the weaker community structure, with increased communication with the
active site (circled). On the other hand, the inactive conformation exhibits
impaired communication with the active site. These images are snapshots from a
video which can be viewed online.}
  \label{fgr:example}
\end{figure*}

\begin{table}
\small
  \caption{Residues with largest conformational ratio $\Delta_{\text{CF}}$ for
random walks originating at residue Glu390.  Residues marked with an asterisk
appear in the hydrogen bonding network identified by Datta \textit{et al}
\cite{datta2008allosteric}.}
  \label{tbl:example}
  \begin{tabular*}{0.5\textwidth}{@{\extracolsep{\fill}}cccc}
    \hline
    Residue &  $t_{1/2}^\text{active}$ & $t_{1/2}^\text{inactive}$ &
$\Delta_{\text{CF}}$ \\
    \hline
    R286\footnotemark[1] & 1 & 493 & 493\\
    G287 & 36 & 729 & 20.25\\
    D288 & 65 & 915 & 14.08\\
    L258 & 2 & 25 & 12.5\\
    S289 & 114 & 1159 & 10.17\\
    S339\footnotemark[1] & 100 & 930 & 9.3\\
    C285 & 36 & 311 & 8.64\\
    S333\footnotemark[1] & 8 & 66 & 8.25\\
    P290 & 193 & 1440 & 7.46\\
    R240 & 244 & 1762 & 7.22\\
    T334 & 18 & 127 & 7.06\\
    G291 & 274 & 1672 & 6.10\\
    W340 & 213 & 1265 & 5.94\\
    \hline
  \end{tabular*}
\end{table}

\textbf{The bond-removal $t_{1/2}$ ratio $\Delta_{\text{BR}}$ identifies bonds
involved in the allosteric hydrogen-bonding network. }
In order to quantify further the contribution of individual bonds to this
communication pathway, we calculate the bond removal ratio $\Delta_{\text{BR}}$ 
for a random walk originating at Glu390 with target at the active site.  Table 3
shows the bonds whose removal induce the largest increases in the characteristic
transient time $t_{1/2}$.  The Glu390/Arg286 salt-bridge has the largest impact,
reflecting the importance of this major pathway. Furthermore, five of the top
six bonds identified in Table 3 are formed within residues known experimentally
to have the greatest impact on catalytic activity~\cite{datta2008allosteric}
(Fig.~7a). 

However, the spread of the random walk shown in Figure 6 and the high importance
of bonds which do not belong to the previously identified allosteric network
points to the existence of minor pathways between the two sites.  In this
respect, we also identify four novel interactions which cause an increase in
$t_{1/2}$ of a similar magnitude: Ser236/Gln283, Arg240/Asp336, Arg341/Thr180,
and Arg286/Asn337 (Fig.~7b). 
 
Arg341 undergoes significant rearrangement between the inactive and active
conformations, transferring from the surface of the protein to the
substrate-binding pocket and conferring selectivity on the substrate through
charge-charge interactions\cite{wilson1994structure}.  Interestingly, it is
conserved across all human caspases. The hydrogen bond it forms with Thr180 may
be important for stabilising its position in the substrate binding pocket. 
Thr179 and Arg341 are important substrate binding residues which provide the
aspartate recognition function of caspase-1.

\begin{figure*}[!t]
\centering
  \includegraphics[width=\textwidth]{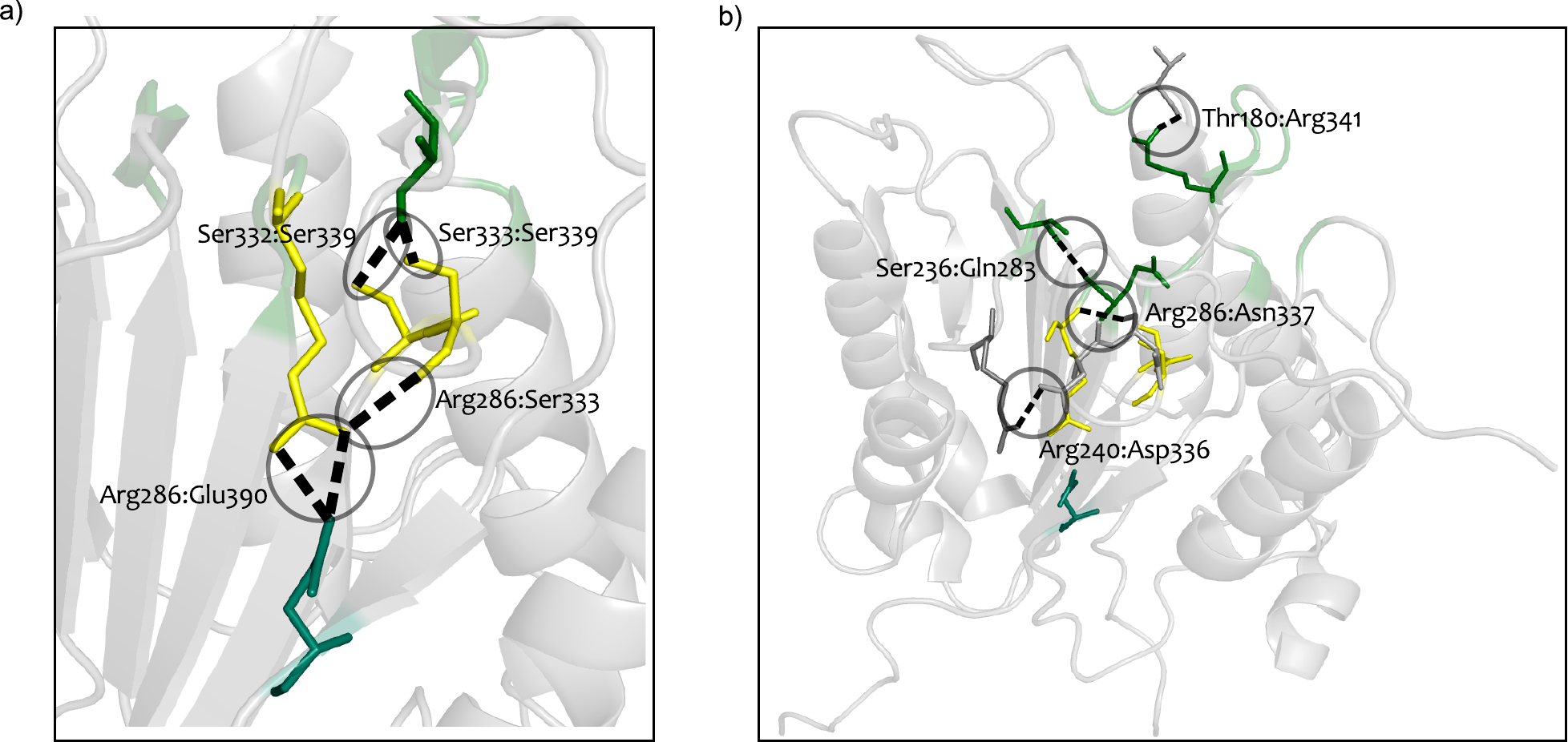}
  \caption{\textbf{Interactions identified by their high bond removal ratio
$\Delta_{\text{BR}}$ as mediating communication between allosteric and active
sites.} In (a), we show bonds with high $\Delta_{\text{BR}}$ which are involved
in the previously identified hydrogen bonding network and are known to have
functional importance.  Previously unidentified interactions, which we suggest
help mediate allosteric communication are shown in (b): $\text{Arg341
H}^{\eta11}:\text{Thr180 O}$,  Arg341 is a crucial residue for substrate
recognition;  $\text{Ser236 H}: \text{Gln283 O}$, this bond weakens in the
active form and the Gln283 sidechain rotates to bind with the substrate;
$\text{Arg240 H}^{\eta11}: \text{Asp336 O}^{\delta1}$, Arg240 is adjacent to
His237 which is believed to form a catalytic diad with Cys285; $\text{Arg286 H}:
\text{Asn337 O}$,  Arg286 is in the L2 loop adjacent to the catalytic residue
Cys285. In both figures, the source of the random walk (residue Glu390) is
coloured blue; the target residues in the active site are coloured green; and
the residues involved in the known hydrogen bonding network are coloured
yellow.}
  \label{fgr:example}
\end{figure*}

Asp336 may stabilise the L3 loop (residues 332-346) in the active
conformation\cite{datta2008allosteric}.  Our analysis here suggests that the
bond it forms with Arg240 is important for maintaining contact with the 236-238
loop containing residue His237 and with Arg286, which is adjacent to the
catalytic residue Cys285.   However, both of these residues were found to have
only a small effect on protein function and so it would seem that these bonds
are not crucial for maintaining the active conformation.  

Gln283 is an active site residue which forms a hydrogen bond with a substrate
sidechain\cite{wilson1994structure}.  Although the Ser236/Gln283 hydrogen bond
is conserved between conformations, it is weaker in the active conformation.  In
the inactive conformation Gln283 forms a hydrogen bond with Ser347 (not
preserved in the active conformation) which serves as an anchor point for the L2
loop containing the catalytic residue Cys285\cite{romanowski2004crystal}.  Thus
the weakening of the Ser236/Gln283 bond could allow Gln283 to rearrange and the
L2 loop to adopt its catalytically competent conformation.  Our analysis
suggests that the Ser236/Gln283 bond lies on a signalling pathway between the
allosteric and active sites and so perturbations induced by binding of the
inhibitor could affect this conformational re-arrangement.  Further experimental
work is required to identify whether Thr180, Gln283, or Ser236 are functionally
significant.
 
\begin{table}
\small
  \caption{Bonds with largest bond removal ratio  $\Delta_{\text{BR}}$ for a
random walk originating at Glu390 towards the active site.  Bonds marked with an
asterisk are involved in the hydrogen bonding network identified by Datta
\textit{et al}\cite{datta2008allosteric}}
  \label{tbl:example}
  \begin{tabular*}{0.5\textwidth}{@{\extracolsep{\fill}}cc}
    \hline
   Bond & $\Delta_{\text{BR}}$ \\
    \hline
    $\text{Arg286} \text{H}^{\eta22}: \text{Glu390}^{\epsilon
1}\footnotemark[1]$ & 1.0693\\
    $\text{Arg286 H}^{\eta12}: \text{Glu390 O}^{\epsilon1}\footnotemark[1]$ &
1.0602\\
    $\text{Ser236 H}: \text{Gln283 O}$ & 1.0545\\
    $\text{Arg286 H}^{\eta11}:  \text{Ser333 O}$\footnotemark[1] & 1.0454\\
    $\text{Ser332 H}^{\gamma} : \text{Ser339 O}^{\gamma}$\footnotemark[1]  &
1.0440\\
    $\text{Ser333}^{\gamma}: \text{Ser339 O}^{\gamma}$\footnotemark[1] &
1.0304\\
    $\text{Arg240 H}^{\eta11}: \text{Asp336 O}^{\delta1}$& 1.0294\\
    $\text{Arg341 H}^{\eta11} : \text{Thr180 O}$ & 1.0264\\
    $\text{Arg286 H}: \text{Asn337 O}$ & 1.0256\\
    \hline
  \end{tabular*}
\end{table}

\subsection{The mutational ratio  $\Delta_{\text{MT}}$ reveals differences
between active-to-allosteric-site and active-to-active-site communication in the
caspase tetramer}

The previous analysis has assumed no information about the location of the
allosteric site, i.e., the location of the allosteric site was only used
\textit{a posteriori} to evaluate the outcomes of our algorithm. If the location
of the allosteric site is known, we can integrate this information in our
analysis of signal propagation. To do this, we use  the mutational $t_{1/2}$
ratio ($\Delta_{\text{MT}}$) defined in Eq.~\eqref{eq:MT}, which measures the
impact of mutating a residue on the communication between two sites.  We have
performed the computational mutagenesis of all residues and calculated their
$\Delta_{\text{MT}}^{\text{Act-Allost}}$ for a random walk between the active
site (defined as above) and the allosteric site (defined as the residues within
3.5\AA ~of the allosteric ligand). The top residues are shown in Table 4.  Four
of the key allosteric residues discussed above are identified as having large
mutational ratios $\Delta_{\text{MT}}^{\text{Act-Allost}}$:  
E390 (1st), R286 (3rd), S339 (7th), and S332 (12th).   
We have also checked our results againts other independent findings. 
In particular, the key allosteric residues previously identified in
Ref.~\citet{datta2008allosteric} (286/332/339/390) have statistically
significant higher $\Delta_{\text{MT}}^{\text{Act-Allost}}$ than other residues
in the protein  (Wilcoxon rank sum test $p=6.5\times10^{-5}$).  Furthermore, 
the residues with highest 
$\Delta_{\text{MT}}^{\text{Act-Allost}}$ correspond with those of highest
functional importance, as shown by the measured impact of the mutation on the
catalytic efficiency $k_{\text{cat}}/K_m$\cite{datta2008allosteric} (Fig.~8c).

In addition to allosteric inhibition, caspase-1 also exhibits strong positive
cooperativity between the two active sites present in its tetramer. Recently, it
has been shown that binding at one active site promotes activity at the other
active-site\cite{datta2013substrate}, possibly due to induced dimerisation or
the propagation of a conformational change.  Interestingly, this cooperative
behaviour is not removed by mutations implicated in allosteric
inhibition\cite{datta2008allosteric}, which suggests that cooperativity is
mediated by a different mechanism. 
We have used our Markov transient analysis to compare the active-to-allosteric
site communication versus the active-to-active site communication related to
cooperativity between the two active sites of the caspase tetramer. To study the
relevant residues involved in cooperative behaviour we have calculated the
mutational $t_{1/2}$ ratios of all residues but now for a random walk
\textit{between the two active sites}, $\Delta^{\text{Act-Act}}_{\text{MT}}$. 
In Figure~8, we show that the $\Delta^{\text{Act-Act}}_{\text{MT}}$ are notably
different to the $\Delta_{\text{MT}}^{\text{Act-Allost}}$ ratios obtained above
for the active-site-to-allosteric-site random walk. For instance, the 
active-to-allosteric mutational ratio  of E390 and R286 drops dramatically,
whilst other residues increase in importance. 

\begin{table}[]
\small
  \caption{Residues with largest mutational ratio for a random walk between
between the active and allosteric sites $\Delta_{\text{MT}}^{\text{Act-Allost}}$
and between the two active sites 
 $\Delta_{\text{MT}}^{\text{Act-Act}}$ in the caspase-1 tetramer.  Residues
marked with an asterisk appear in the hydrogen bonding network identified by
Datta \textit{et al} \cite{datta2008allosteric}.}
  \label{tbl:example}
  \begin{tabular*}{0.5\textwidth}{@{\extracolsep{\fill}}cccc}
    \hline
    \multicolumn{2}{c}{Active-Allosteric}  & \multicolumn{2}{c}{Active-Active}
\\
    \hline
    Residue &  $\Delta_{\text{MT}}^{\text{Act-Allost}}$ & Residue & 
$\Delta_{\text{MT}}^{\text{Act-Act}}$\\
    \hline
 E390* & 1.2590 &  R383 & 1.0845  \\
 C285 & 1.2275 &     E390* & 1.0454  \\
 R286* & 1.2169 &     R286* & 1.0387 \\
 L258 & 1.1584 &     E378 & 1.0338 \\
 I282 & 1.0972 &     S339* & 1.0326  \\
 F262 &  1.0890 & C285 & 1.0288 \\
 S339* & 1.0855 &     T389 & 1.0274 \\
 Q283 & 1.0767 &     R391 & 1.0262 \\ 
 I261 & 1.0737 &     L325 & 1.0253  \\
 I243 & 1.0683 &     N259 & 1.0244 \\
 L256 & 1.0678 &     T334 & 1.0194 \\
 S332* & 1.0669 &     F439 & 1.0169 \\
    \hline
  \end{tabular*}
\end{table}

The residue with largest $\Delta^{\text{Act-Act}}_{\text{MT}}$ for active-site
to active-site communication is Arg383.  In Section~\ref{sec:Markov}, we found
that the strongly compartmental community structure of the inactive conformation
was weakened by mutation of Arg383.  Here we find that this residue is also
important for active-site to active-site communication.  Residues forming
dimer-dimer contacts also play a significant role in communication between the
two active sites.  In addition to Arg383, the residue pairs E378/L325 and
T389/R391, which form weak interactions between the two dimers in the active
conformation, also have high $\Delta^{\text{Act-Act}}_{\text{MT}}$. This
suggests that such connections are important for transmitting signals between
the two active sites and these residues could therefore play a role in
transmitting binding-induced conformational changes.  

\begin{figure}[!t]
\centering
  \includegraphics[]{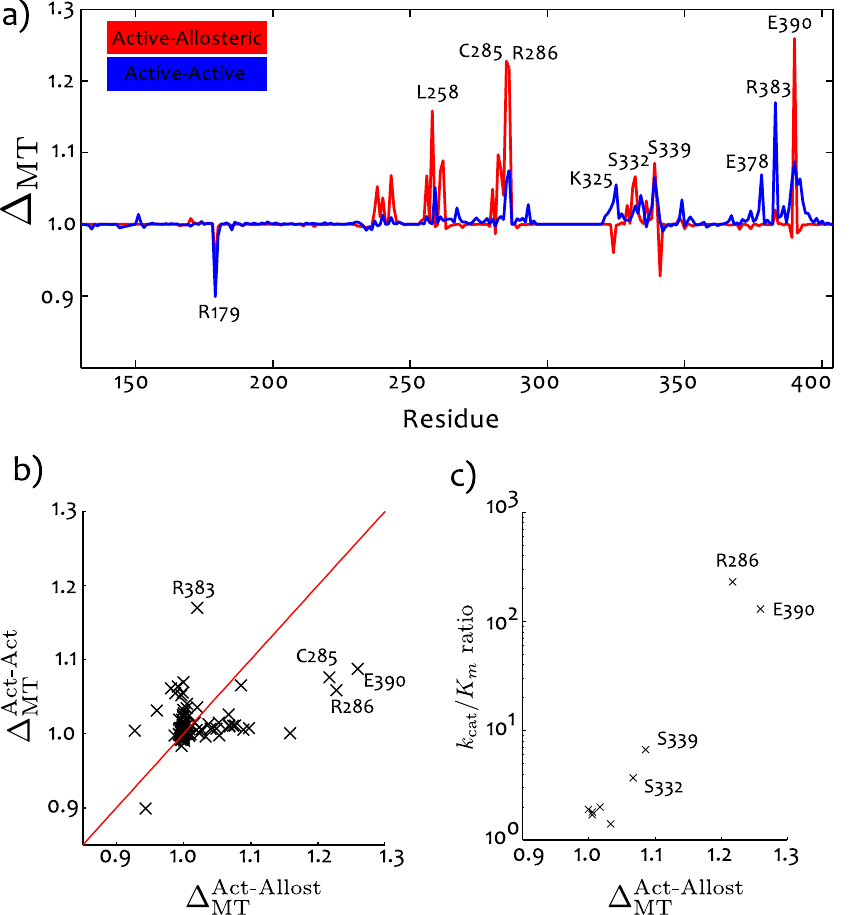}
  \caption{\textbf{Comparing the mutational $t_{1/2}$ ratios between
allosteric-to-active-site and active-to-active-site communication.}  (a) The
mutational $t_{1/2}$ ratio for each residue for random walks between the
active-site and allosteric-site (red) and between the two active-sites (blue)
show differences.  (b) The active-site/active-site ratio
$\Delta^{\text{Act-Act}}_{\text{MT}}$ plotted against the
active-site/allosteric-site ratio $\Delta^{\text{Act-Allost}}_{\text{MT}}$: the
$\Delta^{\text{Act-Act}}_{\text{MT}}$ of residues implicated in allosteric
inhibition (E390/R286) is much lower than their active-site to allosteric-site
$t_{1/2}$ ratio, which suggests that these residues are less important for
co-operative behaviour. (c) $\Delta_{\text{MT}}^{\text{Act-Allost}}$ plotted
against the experimental $k_{\text{cat}}/K_m$ ratio (for residues with available
experimental data\cite{datta2008allosteric}): residues with large
$\Delta_{\text{MT}}^{\text{Act-Allost}}$ correspond to those with greatest
functional significance.}
  \label{fgr:example}
\end{figure}

\section{Conclusions}

In this paper we have used two methods that exploit the transients of a Markov
process diffusing on an atomistic biophysical graph derived from protein
structures in order to study allosteric communication pathways. Firstly, the
Markov Stability community detection method identifies differences in the
multiscale structural organisation of the active and inactive structures of
caspase-1. In contrast to its inactive counterpart, the active conformation
exhibits a fluid, weakly compartmentalised community structure with less robust
partitions at large scales. This suggests that perturbations propagate over
larger distances in the active conformation, due to an increase in long-range
communication pathways.  Computational mutational analysis identifies the bond
between residues Arg383 and Asp336 as crucial for maintaining this modular
organisation. 

Secondly, the analysis of transients of random walks originating in the active
site suggests the existence of long-range communication pathways towards the
allosteric site, which show distinctive characteristics in the active and
inactive conformations.  We have introduced three related quantitative criteria,
the conformational, mutational, and bond-removal $t_{1/2}$ ratios, which allow
us to identify the relevant bonds and residues on these pathways by comparing
the changes in their time-dependent participation in the transient. Using the
conformational $t_{1/2}$ ratio allows us to detect a hot spot at the allosteric
site.  Many of the residues and bonds identified with these criteria correspond
to mutations of known functional significance.  We have also found several novel
interactions which may be involved in alternative allosteric pathways.  Finally,
we have used a full computational point mutagenesis to compare the
allosteric-to-active site communication versus the active-to-active site
communication in the caspase dimer and shown that the relevant residues for
allosteric pathways are distinct from those that play a role in the
communication between active sites related to cooperativity.  This agrees with
experimental findings that mutation of allosteric network residues does not
affect cooperativity.

Our method is a computationally efficient method to study the many parallel
communication pathways in biomolecules in a probabilistic setting. Measures
which go beyond binary comparisons (`present' or
`absent')\cite{datta2008allosteric} of bonds in the active/inactive structures
provide more information about the weak interactions which mediate allosteric
signals.  Our transient random walk approach allows such an analysis and
uncovers bonds lying on multiple pathways between the active and allosteric
sites.  Our Markov transient analysis provides a unified understanding which
brings together structural community detection and random walk pathway
identification, and therefore offers a robust way to look for candidate residues
with important structural and functional roles in proteins.  In particular, this
method can be used to identify residues lying on communication pathways between
an effector site (be it allosteric or another active site) and the active site.

\section{Acknowledgements}

This work was funded by an EPSRC Centre for Doctoral Training Studentship from
the Institute of Chemical Biology (Imperial College London) awarded to BA.

\footnotesize{
\bibliography{caspase} 

\providecommand*{\mcitethebibliography}{\thebibliography}
\csname @ifundefined\endcsname{endmcitethebibliography}
{\let\endmcitethebibliography\endthebibliography}{}
\begin{mcitethebibliography}{45}
\providecommand*{\natexlab}[1]{#1}
\providecommand*{\mciteSetBstSublistMode}[1]{}
\providecommand*{\mciteSetBstMaxWidthForm}[2]{}
\providecommand*{\mciteBstWouldAddEndPuncttrue}
  {\def\EndOfBibitem{\unskip.}}
\providecommand*{\mciteBstWouldAddEndPunctfalse}
  {\let\EndOfBibitem\relax}
\providecommand*{\mciteSetBstMidEndSepPunct}[3]{}
\providecommand*{\mciteSetBstSublistLabelBeginEnd}[3]{}
\providecommand*{\EndOfBibitem}{}
\mciteSetBstSublistMode{f}
\mciteSetBstMaxWidthForm{subitem}
{(\emph{\alph{mcitesubitemcount}})}
\mciteSetBstSublistLabelBeginEnd{\mcitemaxwidthsubitemform\space}
{\relax}{\relax}

\bibitem[Cui and Karplus(2008)]{cui2008allostery}
Q.~Cui and M.~Karplus, \emph{Protein Science}, 2008, \textbf{17},
  1295--1307\relax
\mciteBstWouldAddEndPuncttrue
\mciteSetBstMidEndSepPunct{\mcitedefaultmidpunct}
{\mcitedefaultendpunct}{\mcitedefaultseppunct}\relax
\EndOfBibitem
\bibitem[Koshland~Jr \emph{et~al.}(1966)Koshland~Jr, Nemethy, and
  Filmer]{koshland1966comparison}
D.~Koshland~Jr, G.~Nemethy and D.~Filmer, \emph{Biochemistry}, 1966,
  \textbf{5}, 365--385\relax
\mciteBstWouldAddEndPuncttrue
\mciteSetBstMidEndSepPunct{\mcitedefaultmidpunct}
{\mcitedefaultendpunct}{\mcitedefaultseppunct}\relax
\EndOfBibitem
\bibitem[Monod \emph{et~al.}(1965)Monod, Wyman, and Changeux]{monod1965nature}
J.~Monod, J.~Wyman and J.~Changeux, \emph{Journal of molecular biology}, 1965,
  \textbf{12}, 88--118\relax
\mciteBstWouldAddEndPuncttrue
\mciteSetBstMidEndSepPunct{\mcitedefaultmidpunct}
{\mcitedefaultendpunct}{\mcitedefaultseppunct}\relax
\EndOfBibitem
\bibitem[Swain and Gierasch(2006)]{Gierasch2006}
J.~Swain and L.~Gierasch, \emph{Current Opinion in Structural Biology}, 2006,
  \textbf{16}, 102--108\relax
\mciteBstWouldAddEndPuncttrue
\mciteSetBstMidEndSepPunct{\mcitedefaultmidpunct}
{\mcitedefaultendpunct}{\mcitedefaultseppunct}\relax
\EndOfBibitem
\bibitem[Gunasekaran \emph{et~al.}(2004)Gunasekaran, Ma, and
  Nussinov]{Gunasekaran2004}
K.~Gunasekaran, B.~Ma and R.~Nussinov, \emph{PROTEINS: Structure, Function and
  Bioinformatics}, 2004, \textbf{57}, 433--443\relax
\mciteBstWouldAddEndPuncttrue
\mciteSetBstMidEndSepPunct{\mcitedefaultmidpunct}
{\mcitedefaultendpunct}{\mcitedefaultseppunct}\relax
\EndOfBibitem
\bibitem[Cooper and Dryden(1984)]{cooper1984allostery}
A.~Cooper and D.~Dryden, \emph{European Biophysics Journal}, 1984, \textbf{11},
  103--109\relax
\mciteBstWouldAddEndPuncttrue
\mciteSetBstMidEndSepPunct{\mcitedefaultmidpunct}
{\mcitedefaultendpunct}{\mcitedefaultseppunct}\relax
\EndOfBibitem
\bibitem[Popovych \emph{et~al.}(2006)Popovych, Sun, Ebright, and
  Kalodimos]{popovych2006dynamically}
N.~Popovych, S.~Sun, R.~Ebright and C.~Kalodimos, \emph{Nature structural \&
  molecular biology}, 2006, \textbf{13}, 831--838\relax
\mciteBstWouldAddEndPuncttrue
\mciteSetBstMidEndSepPunct{\mcitedefaultmidpunct}
{\mcitedefaultendpunct}{\mcitedefaultseppunct}\relax
\EndOfBibitem
\bibitem[Henzler-Wildman and Kern(2007)]{henzlerwildman2007}
K.~Henzler-Wildman and D.~Kern, \emph{Nature}, 2007, \textbf{450},
  964--972\relax
\mciteBstWouldAddEndPuncttrue
\mciteSetBstMidEndSepPunct{\mcitedefaultmidpunct}
{\mcitedefaultendpunct}{\mcitedefaultseppunct}\relax
\EndOfBibitem
\bibitem[Zhuravlev \emph{et~al.}(2010)Zhuravlev,
  Papoian,\emph{et~al.}]{zhuravlev2010protein}
P.~Zhuravlev, G.~Papoian \emph{et~al.}, \emph{Quarterly reviews of biophysics},
  2010, \textbf{43}, 295--332\relax
\mciteBstWouldAddEndPuncttrue
\mciteSetBstMidEndSepPunct{\mcitedefaultmidpunct}
{\mcitedefaultendpunct}{\mcitedefaultseppunct}\relax
\EndOfBibitem
\bibitem[del Sol \emph{et~al.}(2009)del Sol, Tsai, Ma, and
  Nussinov]{del2009origin}
A.~del Sol, C.~Tsai, B.~Ma and R.~Nussinov, \emph{Structure}, 2009,
  \textbf{17}, 1042--1050\relax
\mciteBstWouldAddEndPuncttrue
\mciteSetBstMidEndSepPunct{\mcitedefaultmidpunct}
{\mcitedefaultendpunct}{\mcitedefaultseppunct}\relax
\EndOfBibitem
\bibitem[Lockless and Ranganathan(1999)]{lockless1999evolutionarily}
S.~Lockless and R.~Ranganathan, \emph{Science}, 1999, \textbf{286},
  295--299\relax
\mciteBstWouldAddEndPuncttrue
\mciteSetBstMidEndSepPunct{\mcitedefaultmidpunct}
{\mcitedefaultendpunct}{\mcitedefaultseppunct}\relax
\EndOfBibitem
\bibitem[Fuentes \emph{et~al.}(2004)Fuentes, Der, and Lee]{fuentes2004ligand}
E.~Fuentes, C.~Der and A.~Lee, \emph{Journal of molecular biology}, 2004,
  \textbf{335}, 1105--1115\relax
\mciteBstWouldAddEndPuncttrue
\mciteSetBstMidEndSepPunct{\mcitedefaultmidpunct}
{\mcitedefaultendpunct}{\mcitedefaultseppunct}\relax
\EndOfBibitem
\bibitem[Fuentes \emph{et~al.}(2006)Fuentes, Gilmore, Mauldin, and
  Lee]{fuentes2006evaluation}
E.~Fuentes, S.~Gilmore, R.~Mauldin and A.~Lee, \emph{Journal of molecular
  biology}, 2006, \textbf{364}, 337--351\relax
\mciteBstWouldAddEndPuncttrue
\mciteSetBstMidEndSepPunct{\mcitedefaultmidpunct}
{\mcitedefaultendpunct}{\mcitedefaultseppunct}\relax
\EndOfBibitem
\bibitem[Kong and Karplus(2009)]{kong2009signaling}
Y.~Kong and M.~Karplus, \emph{Proteins: Structure, Function, and
  Bioinformatics}, 2009, \textbf{74}, 145--154\relax
\mciteBstWouldAddEndPuncttrue
\mciteSetBstMidEndSepPunct{\mcitedefaultmidpunct}
{\mcitedefaultendpunct}{\mcitedefaultseppunct}\relax
\EndOfBibitem
\bibitem[Ota and Agard(2005)]{ota2005intramolecular}
N.~Ota and D.~Agard, \emph{Journal of molecular biology}, 2005, \textbf{351},
  345--354\relax
\mciteBstWouldAddEndPuncttrue
\mciteSetBstMidEndSepPunct{\mcitedefaultmidpunct}
{\mcitedefaultendpunct}{\mcitedefaultseppunct}\relax
\EndOfBibitem
\bibitem[Hilser \emph{et~al.}(2006)Hilser, E., Oas, Kapp, and
  Whitten]{hilser2006statistical}
V.~Hilser, B.~G.-M. E., T.~Oas, G.~Kapp and S.~Whitten, \emph{Chemical
  reviews}, 2006, \textbf{106}, 1545--1558\relax
\mciteBstWouldAddEndPuncttrue
\mciteSetBstMidEndSepPunct{\mcitedefaultmidpunct}
{\mcitedefaultendpunct}{\mcitedefaultseppunct}\relax
\EndOfBibitem
\bibitem[Pan \emph{et~al.}(2000)Pan, Lee, and Hilser]{pan2000binding}
H.~Pan, J.~Lee and V.~Hilser, \emph{Proceedings of the National Academy of
  Sciences}, 2000, \textbf{97}, 12020\relax
\mciteBstWouldAddEndPuncttrue
\mciteSetBstMidEndSepPunct{\mcitedefaultmidpunct}
{\mcitedefaultendpunct}{\mcitedefaultseppunct}\relax
\EndOfBibitem
\bibitem[Del~Sol \emph{et~al.}(2006)Del~Sol, Fujihashi, Amoros, and
  Nussinov]{del2006residues}
A.~Del~Sol, H.~Fujihashi, D.~Amoros and R.~Nussinov, \emph{Molecular systems
  biology}, 2006, \textbf{2}, n/a\relax
\mciteBstWouldAddEndPuncttrue
\mciteSetBstMidEndSepPunct{\mcitedefaultmidpunct}
{\mcitedefaultendpunct}{\mcitedefaultseppunct}\relax
\EndOfBibitem
\bibitem[Del~Sol \emph{et~al.}(2007)Del~Sol, Ara{\'u}zo-Bravo, Amoros,
  Nussinov,\emph{et~al.}]{del2007modular}
A.~Del~Sol, M.~Ara{\'u}zo-Bravo, D.~Amoros, R.~Nussinov \emph{et~al.},
  \emph{Genome Biol}, 2007, \textbf{8}, R92\relax
\mciteBstWouldAddEndPuncttrue
\mciteSetBstMidEndSepPunct{\mcitedefaultmidpunct}
{\mcitedefaultendpunct}{\mcitedefaultseppunct}\relax
\EndOfBibitem
\bibitem[Sethi \emph{et~al.}(2009)Sethi, Eargle, Black, and
  Luthey-Schulten]{sethi2009dynamical}
A.~Sethi, J.~Eargle, A.~Black and Z.~Luthey-Schulten, \emph{Proceedings of the
  National Academy of Sciences}, 2009, \textbf{106}, 6620\relax
\mciteBstWouldAddEndPuncttrue
\mciteSetBstMidEndSepPunct{\mcitedefaultmidpunct}
{\mcitedefaultendpunct}{\mcitedefaultseppunct}\relax
\EndOfBibitem
\bibitem[Ghosh and Vishveshwara(2007)]{ghosh2007study}
A.~Ghosh and S.~Vishveshwara, \emph{Proceedings of the National Academy of
  Sciences}, 2007, \textbf{104}, 15711\relax
\mciteBstWouldAddEndPuncttrue
\mciteSetBstMidEndSepPunct{\mcitedefaultmidpunct}
{\mcitedefaultendpunct}{\mcitedefaultseppunct}\relax
\EndOfBibitem
\bibitem[Chennubhotla and Bahar(2007)]{chennubhotla2007signal}
C.~Chennubhotla and I.~Bahar, \emph{PLoS computational biology}, 2007,
  \textbf{3}, e172\relax
\mciteBstWouldAddEndPuncttrue
\mciteSetBstMidEndSepPunct{\mcitedefaultmidpunct}
{\mcitedefaultendpunct}{\mcitedefaultseppunct}\relax
\EndOfBibitem
\bibitem[Park and Kim(2011)]{park2011modeling}
K.~Park and D.~Kim, \emph{BMC bioinformatics}, 2011, \textbf{12}, S23\relax
\mciteBstWouldAddEndPuncttrue
\mciteSetBstMidEndSepPunct{\mcitedefaultmidpunct}
{\mcitedefaultendpunct}{\mcitedefaultseppunct}\relax
\EndOfBibitem
\bibitem[Lu and Liang(2009)]{lu2009perturbation}
H.~Lu and J.~Liang, \emph{PLoS Computational Biology}, 2009, \textbf{5},
  e1000526\relax
\mciteBstWouldAddEndPuncttrue
\mciteSetBstMidEndSepPunct{\mcitedefaultmidpunct}
{\mcitedefaultendpunct}{\mcitedefaultseppunct}\relax
\EndOfBibitem
\bibitem[Chennubhotla and Bahar(2006)]{chennubhotla2006markov}
C.~Chennubhotla and I.~Bahar, \emph{Molecular systems biology}, 2006,
  \textbf{2}, n/a\relax
\mciteBstWouldAddEndPuncttrue
\mciteSetBstMidEndSepPunct{\mcitedefaultmidpunct}
{\mcitedefaultendpunct}{\mcitedefaultseppunct}\relax
\EndOfBibitem
\bibitem[Delvenne \emph{et~al.}(2010)Delvenne, Yaliraki, and
  Barahona]{delvenne2010stability}
J.~Delvenne, S.~Yaliraki and M.~Barahona, \emph{Proceedings of the National
  Academy of Sciences}, 2010, \textbf{107}, 12755--12760\relax
\mciteBstWouldAddEndPuncttrue
\mciteSetBstMidEndSepPunct{\mcitedefaultmidpunct}
{\mcitedefaultendpunct}{\mcitedefaultseppunct}\relax
\EndOfBibitem
\bibitem[Newman(2006)]{newman2006modularity}
M.~E. Newman, \emph{Proceedings of the National Academy of Sciences}, 2006,
  \textbf{103}, 8577--8582\relax
\mciteBstWouldAddEndPuncttrue
\mciteSetBstMidEndSepPunct{\mcitedefaultmidpunct}
{\mcitedefaultendpunct}{\mcitedefaultseppunct}\relax
\EndOfBibitem
\bibitem[Fortunato and Barthelemy(2007)]{fortunato2007resolution}
S.~Fortunato and M.~Barthelemy, \emph{Proceedings of the National Academy of
  Sciences}, 2007, \textbf{104}, 36--41\relax
\mciteBstWouldAddEndPuncttrue
\mciteSetBstMidEndSepPunct{\mcitedefaultmidpunct}
{\mcitedefaultendpunct}{\mcitedefaultseppunct}\relax
\EndOfBibitem
\bibitem[Schaub \emph{et~al.}(2012)Schaub, Delvenne, Yaliraki, and
  Barahona]{schaub2012markov}
M.~T. Schaub, J.-C. Delvenne, S.~N. Yaliraki and M.~Barahona, \emph{PloS one},
  2012, \textbf{7}, e32210\relax
\mciteBstWouldAddEndPuncttrue
\mciteSetBstMidEndSepPunct{\mcitedefaultmidpunct}
{\mcitedefaultendpunct}{\mcitedefaultseppunct}\relax
\EndOfBibitem
\bibitem[Scheer \emph{et~al.}(2006)Scheer, Romanowski, and
  Wells]{scheer2006common}
J.~Scheer, M.~Romanowski and J.~Wells, \emph{Proceedings of the National
  Academy of Sciences}, 2006, \textbf{103}, 7595--7600\relax
\mciteBstWouldAddEndPuncttrue
\mciteSetBstMidEndSepPunct{\mcitedefaultmidpunct}
{\mcitedefaultendpunct}{\mcitedefaultseppunct}\relax
\EndOfBibitem
\bibitem[Datta \emph{et~al.}(2008)Datta, Scheer, Romanowski, and
  Wells]{datta2008allosteric}
D.~Datta, J.~Scheer, M.~Romanowski and J.~Wells, \emph{Journal of molecular
  biology}, 2008, \textbf{381}, 1157--1167\relax
\mciteBstWouldAddEndPuncttrue
\mciteSetBstMidEndSepPunct{\mcitedefaultmidpunct}
{\mcitedefaultendpunct}{\mcitedefaultseppunct}\relax
\EndOfBibitem
\bibitem[Sleath \emph{et~al.}(1990)Sleath, Hendrickson, Kronheim, March, and
  Black]{sleath1990substrate}
P.~Sleath, R.~Hendrickson, S.~Kronheim, C.~March and R.~Black, \emph{Journal of
  Biological Chemistry}, 1990, \textbf{265}, 14526--14528\relax
\mciteBstWouldAddEndPuncttrue
\mciteSetBstMidEndSepPunct{\mcitedefaultmidpunct}
{\mcitedefaultendpunct}{\mcitedefaultseppunct}\relax
\EndOfBibitem
\bibitem[Li and Yuan(2008)]{li2008caspases}
J.~Li and J.~Yuan, \emph{Oncogene}, 2008, \textbf{27}, 6194--6206\relax
\mciteBstWouldAddEndPuncttrue
\mciteSetBstMidEndSepPunct{\mcitedefaultmidpunct}
{\mcitedefaultendpunct}{\mcitedefaultseppunct}\relax
\EndOfBibitem
\bibitem[Wilson \emph{et~al.}(1994)Wilson, Black, Thomson, Kim, Griffith,
  Navia, Murcko, Chambers, Aldape, Raybuck,\emph{et~al.}]{wilson1994structure}
K.~Wilson, J.~Black, J.~Thomson, E.~Kim, J.~Griffith, M.~Navia, M.~Murcko,
  S.~Chambers, R.~Aldape, S.~Raybuck \emph{et~al.}, \emph{Nature}, 1994,
  \textbf{370}, 270--275\relax
\mciteBstWouldAddEndPuncttrue
\mciteSetBstMidEndSepPunct{\mcitedefaultmidpunct}
{\mcitedefaultendpunct}{\mcitedefaultseppunct}\relax
\EndOfBibitem
\bibitem[Romanowski \emph{et~al.}(2004)Romanowski, Scheer, O'Brien, and
  McDowell]{romanowski2004crystal}
M.~Romanowski, J.~Scheer, T.~O'Brien and R.~McDowell, \emph{Structure}, 2004,
  \textbf{12}, 1361--1371\relax
\mciteBstWouldAddEndPuncttrue
\mciteSetBstMidEndSepPunct{\mcitedefaultmidpunct}
{\mcitedefaultendpunct}{\mcitedefaultseppunct}\relax
\EndOfBibitem
\bibitem[Berman \emph{et~al.}(2000)Berman, Westbrook, Feng, Gilliland, Bhat,
  Weissig, Shindyalov, and Bourne]{berman2000protein}
H.~Berman, J.~Westbrook, Z.~Feng, G.~Gilliland, T.~Bhat, H.~Weissig,
  I.~Shindyalov and P.~Bourne, \emph{Nucleic acids research}, 2000,
  \textbf{28}, 235--242\relax
\mciteBstWouldAddEndPuncttrue
\mciteSetBstMidEndSepPunct{\mcitedefaultmidpunct}
{\mcitedefaultendpunct}{\mcitedefaultseppunct}\relax
\EndOfBibitem
\bibitem[Delmotte \emph{et~al.}(2011)Delmotte, Tate, Yaliraki, and
  Barahona]{delmotte2011protein}
A.~Delmotte, E.~Tate, S.~Yaliraki and M.~Barahona, \emph{Physical Biology},
  2011, \textbf{8}, 055010\relax
\mciteBstWouldAddEndPuncttrue
\mciteSetBstMidEndSepPunct{\mcitedefaultmidpunct}
{\mcitedefaultendpunct}{\mcitedefaultseppunct}\relax
\EndOfBibitem
\bibitem[Word \emph{et~al.}(1999)Word, Lovell, Richardson, and
  Richardson]{word1999asparagine}
J.~M. Word, S.~C. Lovell, J.~S. Richardson and D.~C. Richardson, \emph{Journal
  of molecular biology}, 1999, \textbf{285}, 1735--1747\relax
\mciteBstWouldAddEndPuncttrue
\mciteSetBstMidEndSepPunct{\mcitedefaultmidpunct}
{\mcitedefaultendpunct}{\mcitedefaultseppunct}\relax
\EndOfBibitem
\bibitem[Jacobs \emph{et~al.}(2001)Jacobs, Rader, Kuhn, and
  Thorpe]{jacobs2001protein}
D.~Jacobs, A.~Rader, L.~Kuhn and M.~Thorpe, \emph{Proteins: Structure,
  Function, and Bioinformatics}, 2001, \textbf{44}, 150--165\relax
\mciteBstWouldAddEndPuncttrue
\mciteSetBstMidEndSepPunct{\mcitedefaultmidpunct}
{\mcitedefaultendpunct}{\mcitedefaultseppunct}\relax
\EndOfBibitem
\bibitem[Mayo \emph{et~al.}(1990)Mayo, Olafson, and Goddard]{mayo1990dreiding}
S.~Mayo, B.~Olafson and W.~Goddard, \emph{Journal of Physical Chemistry}, 1990,
  \textbf{94}, 8897--8909\relax
\mciteBstWouldAddEndPuncttrue
\mciteSetBstMidEndSepPunct{\mcitedefaultmidpunct}
{\mcitedefaultendpunct}{\mcitedefaultseppunct}\relax
\EndOfBibitem
\bibitem[Blondel \emph{et~al.}(2008)Blondel, Guillaume, Lambiotte, and
  Lefebvre]{blondel2008fast}
V.~Blondel, J.~Guillaume, R.~Lambiotte and E.~Lefebvre, \emph{Journal of
  Statistical Mechanics: Theory and Experiment}, 2008, \textbf{2008},
  P10008\relax
\mciteBstWouldAddEndPuncttrue
\mciteSetBstMidEndSepPunct{\mcitedefaultmidpunct}
{\mcitedefaultendpunct}{\mcitedefaultseppunct}\relax
\EndOfBibitem
\bibitem[Karrer \emph{et~al.}(2008)Karrer, Levina, and
  Newman]{karrer2008robustness}
B.~Karrer, E.~Levina and M.~Newman, \emph{Physical Review E}, 2008,
  \textbf{77}, 046119\relax
\mciteBstWouldAddEndPuncttrue
\mciteSetBstMidEndSepPunct{\mcitedefaultmidpunct}
{\mcitedefaultendpunct}{\mcitedefaultseppunct}\relax
\EndOfBibitem
\bibitem[Meil{\u{a}}(2007)]{meilua2007comparing}
M.~Meil{\u{a}}, \emph{Journal of Multivariate Analysis}, 2007, \textbf{98},
  873--895\relax
\mciteBstWouldAddEndPuncttrue
\mciteSetBstMidEndSepPunct{\mcitedefaultmidpunct}
{\mcitedefaultendpunct}{\mcitedefaultseppunct}\relax
\EndOfBibitem
\bibitem[Rasmussen(2006)]{rasmussen2006gaussian}
C.~E. Rasmussen, \emph{Gaussian processes for machine learning}, MIT Press,
  2006\relax
\mciteBstWouldAddEndPuncttrue
\mciteSetBstMidEndSepPunct{\mcitedefaultmidpunct}
{\mcitedefaultendpunct}{\mcitedefaultseppunct}\relax
\EndOfBibitem
\bibitem[Datta \emph{et~al.}(2013)Datta, McClendon, Jacobson, and
  Wells]{datta2013substrate}
D.~Datta, C.~L. McClendon, M.~P. Jacobson and J.~A. Wells, \emph{Journal of
  Biological Chemistry}, 2013, \textbf{288}, 9971--9981\relax
\mciteBstWouldAddEndPuncttrue
\mciteSetBstMidEndSepPunct{\mcitedefaultmidpunct}
{\mcitedefaultendpunct}{\mcitedefaultseppunct}\relax
\EndOfBibitem
\end{mcitethebibliography}
\bibliographystyle{rsc} 
}

\end{document}